\documentclass{aa}

\usepackage{graphicx, epstopdf, multicol, blindtext, isotope,mathptmx}
\usepackage{txfonts, color, caption, subcaption, paralist, longtable}
\usepackage[normalem]{ulem}
\usepackage{natbib}
\bibpunct{(}{)}{;}{a}{}{,}
\begin{document}

   \title{Velocity limits in the thermonuclear supernova ejection scenario for hypervelocity stars and the origin of US 708}

   \author{P. Neunteufel
          \inst{1,2}}

   \institute{\inst{1}Max Planck Institut f\"ur Astrophysik,
              Karl-Schwarzschild-Straße 1, 85748 Garching bei M\"unchen\\
  \inst{2}University of Leicester,
   	University Road, LE1 7RH Leicester, Leicestershire \\
   \email{pneun@mpa-garching.mpg.de} }

   \date{Received (month) (day), (year); accepted (month) (day), (year)}
\abstract
   {Hypervelocity stars (HVS) are a class of stars moving at high enough velocities to be gravitationally unbound from the Galaxy. In recent years, ejection from a close binary system in which one of the components undergoes a thermonuclear supernova (SN) has emerged as a promising candidate production mechanism for the least massive specimens of this class. The explosion mechanisms leading to thermonuclear supernovae, which include the important Type Ia, and related subtypes, remain unclear. }
   {This study presents a thorough theoretical analysis of candidate progenitor systems of thermonuclear SNe in the single degenerate helium donor scenario in the relevant parameter space leading to the ejection of HVS. The primary goal is investigation of the, previously unclear, characteristics of the velocity spectra of the ejected component, including possible maxima and minima, constraints arising from stellar evolution and initial masses. Further, the question of whether knowledge of the ejection velocity spectra may aid in reconstruction of the terminal state of the supernova progenitor, is addressed, with a focus on the observed object US 708.}
   {Presented are the results of 390 binary model sequences computed with the Modules for Experiments in Stellar Astrophysics (MESA) framework, investigating the evolution of supernova progenitors composed of a helium-rich hot subdwarf and a accreting white dwarf. Assumption of a specific explosion mechanism is avoided as far as possible. The detailed evolution of the donor star as well as gravitational wave radiation and mass transfer driven orbital evolution are fully taken into account. Results are then correlated with an idealized kinematic analysis of the observed object US 708.}
   {It is seen that the ejection velocity spectra reach a maximum in the range $0.19~M_\odot < M_{HVS} < 0.25~M_\odot$. Depending on the local Galactic potential, all donors below $0.4~\text{M}_\odot$ are expected to become HVS. The single degenerate helium donor channel is able to account for runaway velocities up to $\sim1150~\text{km\,s}^{-1}$ with a Chandrasekhar mass accretor, exceeding $1200~\text{km\,s}^{-1}$ if super-Chandrasekhar mass detonations are taken into account. It is found that the previously assumed mass of $0.3~M_\odot$ for US 708, combined with more recently obtained proper motions, favor a sub-Chandrasekhar mass explosion with a terminal WD mass between $1.1~M_\odot$ and $1.2~M_\odot$, while a Chandrasekhar mass explosion requires a mass of $>0.34~M_\odot$ for US 708. This mechanism may be a source of isolated runaway extremely low mass white dwarfs.}
  {The presence of clear, terminal accretor mass dependent, but initial-condition independent, ejection velocity maxima, provides constraints on the terminal state of a supernova progenitor. Depending on the accuracy of astrometry, it is possible to discern certain types of explosion mechanisms from the inferred ejection velocities alone, with current proper motions allowing for a sub- Chandrasekhar mass SN as an origin of US 708. However, more robust reconstructions of the most likely SN progenitor state will require a greater number of observed objects than are currently available.  }
   
   \titlerunning{Hypervelocity star ejection}
\maketitle

\section{Introduction} \label{sec:introduction}

The existence of stars moving at velocities high enough to be unbound from the Galaxy, known as hypervelocity stars (HVS), was first proposed more than three decades ago by \cite{H1988}. While this initial prediction was followed up by a number of theoretical studies \citep{H1991,H1992}, primarily based on the assumption that these object would result from interaction of a star or binary star with a \citep{YT2003}, massive black hole (MBH), observational evidence was not forthcoming.

This changed with the discovery of SDSS J090745.0+024507 by \cite{BGK2005}. Found in the Sloan Digital Sky Survey data set, this object, also referred to as HVS1, is likely a B-type star with a mass of about $3~\text{M}_\odot$ at a distance of $111~\text{kpc}$ from the Galactic center and a Galactic rest frame velocity of $696~\text{km\,s}^{-1}$ \citep{BGK2007}. This object led to a targeted search, conducted by \cite{BGK2007}, which yielded nine further objects in short order, growing to 23 objects within the following decade \citep{B2015}.

Theoretical modeling of the ejection mechanism at this point in time focused heavily on the black hole connection. \cite{BGB2006} predicted the likely ejection velocity spectrum of HVS produced by the encounter of a binary star with a super massive black hole (SMBH) at the Galactic center ($\text{Sag}~A^\ast$). \cite{LL2008} pointed out that an encounter with a stellar mass black hole in orbit around $\text{Sag}~A^\ast$ could also act as a source of HVS with \cite{SMH2009} coming to the same conclusion regarding intermediate mass black holes (IMBH). The latter mechanism has been suggested as a source of certain B-type HVS \citep{IGH2019}. Very recently, discovery of an A-type star of $2.35~\text{M}_\odot$, moving at a Galactic rest frame velocity of $\sim1700~\text{km\,s}^{-1}$, and which could be traced with high accuracy back to the Galactic center \citep{KBD2020}, has greatly bolstered the viability of the SMBH-encounter scenario. For a review of the state of the art on this ejection mechanism, the reader is directed to \cite{B2015}.

Alternative ejection mechanisms to black hole interaction started being considered in 2009 when \cite{ANS2009} proposed, based on the spatial distribution of known HVS, an origin in multiple body interaction during the passing of a satellite dwarf galaxy through the Milky Way. The proposal by \cite{JWP2009} that some HVS could be former members of close binary system with a white dwarf (WD) companion undergoing a supernova (SN) explosion provided a conceptual bridge to ejection mechanisms earlier proposed for runaway stars (RS) gravitationally bound to the Galaxy \citep{B1961,HBZ2001}, earlier considered to be incapable of providing the necessary ejection velocities. This idea was later followed up theoretically for both B and G/K-dwarf type stars in core collapse SNe \citep{T2015T} and hot subdwarfs in the thermonuclear scenario \citep{NYL2016,NYL2019}. 

Observational evidence for the viability of the supernova ejection scenario came when \cite{GFZ2015} pointed out that the, previously known \citep{HHOB2005}, helium sdO, of N-type, indicating unusually fast rotation, US 708 (HVS2) was moving with a greater velocity than originally reported and could not be traced back to the Galactic center.

More recently, accompanying observational successes \citep[e.g.][]{RHK2019}, theoretical interest has focused on the evolution of an SN-ejected HVS in the phase after the supernova explosion \citep{ZF2019,BWB2019}, both under the assumption of the HVS being the former donor star and or a partially burnt remnant of the former accretor.

With the advent of the latest generation of large scale astrometric surveys \citep{GAIA-DR2}, a number of new objects \citep{SBG2018}, thought to originate from a thermonuclear supernova occurring in a double WD system, as well as a number of additional candidates \citep{FF2019}, have been discovered. In the same vein, data obtained by the Gaia instrument has confirmed the origin of HVS3 (HE 0437-5439) in the Large Magellanic cloud \citep{ENH2005,IKH2018,EBG2019}. Considering the capabilities of upcoming instruments \citep[e.g. 4MOST:][]{4MOST1}, further discoveries in this field may be expected.

The supernova ejection scenario for HVS can be considered of particular attractiveness, since, as will be shown, the ejection velocity spectrum for these objects is closely related to the progenitor binary's orbital parameters and mass distribution at the point of HVS ejection. Knowledge of these parameters, which contain information of the state of the exploding companion, can then be used to infer the explosion mechanism of the supernova, which, in the case of thermonuclear events, is still not conclusively resolved \citep[see, e.g., review by][]{HN2000}.

Studies of the pre-explosion evolution of close binary systems have been performed in the past \citep[e.g. ][]{EF1990,YL2004b,YL2004a,Yu2008,WJH2013,NYL2016,NYL2019}, with a focus not on runaway velocities but on the ignition behavior of the accretor. However, while \cite{H2008} and \cite{WH2009} did study the ejection velocity distribution of donor stars ejected from these systems subsequent to a supernova explosion using a population synthesis framework, their parameter space is limited to runaway masses $\geq 0.6~\text{M}_\odot$. The present study is meant to remedy this situation, presenting a detailed examination of the parameter space available for close white dwarf binary systems, including initial mass and mass ratio, as well as initial orbital separation, proposed to allow for the occurrence of thermonuclear SNe. As the precise explosion mechanism of these events is currently still unresolved, the methodology aims to remain agnostic to it and, therefore, to the precise state of the accretor at the time of creation of the HVS. This paper presents the results of 390 detailed binary evolution models sequences, commenting on the ejection velocity spectra of a variety of proposed explosion mechanisms of thermonuclear SNe, the pre-explosion evolution of the system and the donor star and the viability of using HVS as probes of thermonuclear SN explosion mechanisms.

It should be emphasized that the focus of this study is HVS ejection by thermonuclear SNe in general, of which Type Ia SNe are considered a subtype, not Type Ia SNe exclusively, whose relatively similar peak luminosity hints at a likewise relatively similar terminal WD mass at the point of explosion \citep{P1993}. However, in the absence of a consensus on the spectral classification of hypothetical progenitor SNe of observed HVS, it is reasonable to assume that WD masses at the point of detonation in these events may be dissimilar to those responsible for Type Ia SNe. This paper can make no statement on the observational properties of the SN event, only on the velocity of the expected runaway for a certain assumed WD mass.

This report is organized as follows: Sec.~\ref{sec:expl} presents a brief review of the literature concerning explosion mechanisms of thermonuclear SNe. In Sec.~\ref{sec:assumption}, a number of analytical considerations relevant for the investigation of close binary systems are presented. Sec.~\ref{sec:methods} comments on the numerical tools used in this project and justifies the choice of initial model parameters. Sec.~\ref{sec:results} presents the study's findings, commenting on the bulk properties of the sample and observational properties of certain individual cases. Sec.~\ref{sec:US708} presents a simple application to the observed object US 708, of the predictions in the preceding sections. In Sec.~\ref{sec:discussion}, results are discussed, with a brief summary and conclusions in Sec.~\ref{sec:conclusions}. A brief investigation of the effects of variations in the initial orbital separation is shown in Appendix~\ref{app:as}. Sec.~\ref{sec:US708} makes extensive use of calculations of the motion of a hypervelocity star in the Galactic potential. As this is slightly outside the focus of this study, the potential and numerical tool employed here are briefly discussed in Appendix~\ref{app:traj}.

\section{Review of applicable explosion mechanisms} \label{sec:expl}

While it is largely accepted that thermonuclear SNe result from the thermonuclear detonation of a WD receiving material from a binary companion, the nature of the progenitor system is currently not well understood. Generally, hypotheses regarding the companion fall into two distinct categories: Double degenerate (DD), where the companion in another WD, and single degenerate (SD), where the companion is a non-degenerate star (e.g. a main sequence star)\footnote{Note that for our purposes the defining characteristic for categorization into these groups consists solely in the interaction between either two degenerate objects or one degenerate and one non-degenerate object \citep[compare][]{S2013}.}.
However, apart from the reasonably well established identity of the exploding object as a WD, uncertainties persist as to its state, especially its mass, at the point of detonation (terminal mass) as well as the outcome of the explosion. 
It is classically understood that in systems containing two WDs of sufficient mass, Roche lobe overflow (RLOF) leads to one or both companions being dynamically disrupted before merging, with the merged object detonating \citep[see e.g][]{W1984}. While it is generally expected that this scenario will leave no bound remnant, certain violent merger scenarios have been predicted to occur on short enough a timescale to allow for the ejection of a bound object \citep{PKT2013,SBG2018}. Note that dynamical disruption of the mass donor can be avoided in double degenerate systems with relatively small ($q \leq 0.63$) mass ratios \citep{E2011B}. However, as we are concerned with the production of non-degenerate HVS, the DD scenario will be disregarded.
In the SD-scenario, the binary is not expected to merge. Instead, the non-degenerate companion will donate material to the WD (note that this mass transfer is, depending on the mass ratio of the system and the evolutionary state of the donor, not necessarily stable). Depending on the prevalent explosion mechanism, which, as of this time, is still heavily debated, a thermonuclear explosion is initiated on the WD by one of the mechanisms discussed below as soon as the requisite ignition conditions are reached. If the WD is completely disrupted by the explosion, the donor star is, under preservation of its orbital angular momentum, flung away from the former location of the binary with a velocity slightly greater than its terminal orbital velocity \citep{BWB2019}. 
It should be noted that not all hypotheses concerning the mechanism of thermonuclear SNe predict the complete disruption of the WD, instead leaving a partially burnt remnant \citep{VNK2017}. A number of objects fitting this scenario have recently been described \citep{RHK2019}. This would imply that the runaway velocity will be lower than in the case of a complete disruption of the WD (with a possibility of the partially burnt remnant becoming a HVS). While doubtlessly important, closer study of this case is beyond the scope of this paper and is left for future inquiry.
As will be further commented on in Sec.~\ref{sec:assumption}, obtainable velocities in the SD supernova ejection mechanism are inversely correlated with the physical radius of the donor star. This circumstance suggests that helium-rich donor stars in their core helium burning phase are the principal candidates for the production of non-degenerate HVS in the supernova ejection scenario. 
This study therefore chiefly considers production mechanisms for thermonuclear SNe relying on accretion from non-degenerate He-rich donor stars in their core helium burning phase and assumes that the explosion of the WD will leave no appreciable bound remnant.

\begin{figure} 
	\input{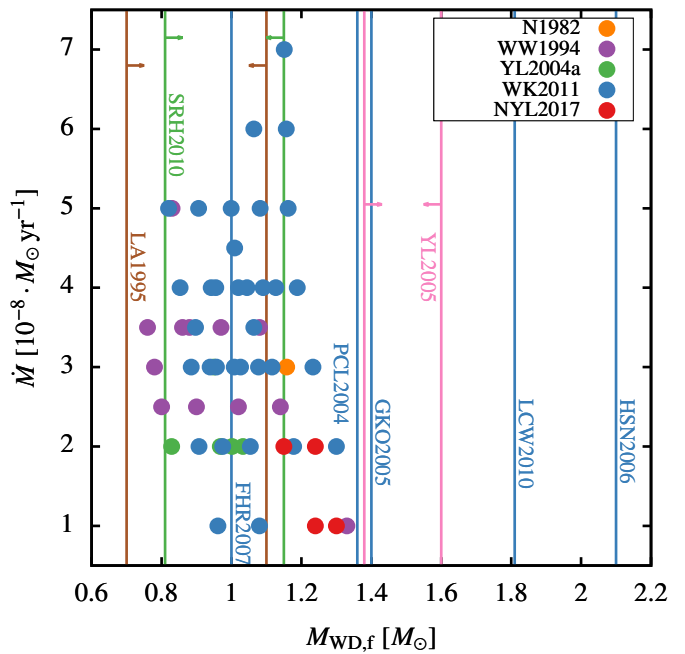}
	\caption{Representation of the parameters of proposed explosion mechanisms in the parameter space accessible to this study with $\dot{M}$ the mass transfer rate and $M_\text{WD,f}$ the proposed total mass of the accreting WD at the point of detonation. If both values are provided by the source, each model is represented as a point. If the source only provides a single value for $M_\text{WD,f}$, the model is represented as a blue line. If a range for $M_\text{WD,f}$ is given, it is represented as pairs of colored lines with arrows indicating the position of the corresponding second line. Labels indicate source material as defined in Table~\ref{table:sources}. } \label{fig:scenarios}
\end{figure}

Keeping this in mind, as well as the importance of the terminal mass of the WD to the runaway velocity, we order our given test cases into three broad categories: Chandrasekhar mass mechanisms, sub-Chandrasekhar mass mechanisms and super-Chandrasekhar mass mechanisms.
The following discussion is conveniently summarized in Fig.~\ref{fig:scenarios} and related citations in Tab.~\ref{table:sources}, showing likely terminal WD masses as proposed in literature at the time of writing.

\subsection{Chandrasekhar mass mechanisms}

Mechanisms falling into this category are distinguished by the assumption that a thermonuclear SN is initiated when the accreting WD reaches a terminal mass close to the Chandrasekhar mass \citep{HF1960, A1969, HW1969}.
While a number of successes in the spectral modeling of explosion mechanisms in this category could be achieved \citep[see e.g.][]{NTY1984, HN2000}, major challenges persist in resolving the dominant one. 
Prompt detonation at the time terminal mass is attained is generally ruled out \citep{A1971} on the grounds of significant overproduction of iron group elements (IGE), while pure deflagration (i.e. subsonic flame propagation, as opposed to supersonic flame propagation in a detonation) models tend to produce insufficient amounts of IGE in addition to featuring insufficiently high ejecta velocities.

A method to overcome this dichotomy was proposed in the deflagration-detonation scenario (DDT), which presupposes ignition of core carbon burning in the subsonic regime, transitioning to the supersonic regime during the time the burning front takes to traverse the WD \citep{KOW1997}. While this scenario generally produces isotope yields well in agreement with observation, major challenges persist in the self-consistent modeling of the subsonic-supersonic transition, as well as that of the microphysics involved in the propagation of the burning front.

Further uncertainties persist in whether the deflagration, once initiated, develops into a detonation after an initial delay (the delayed DDT scenario) - prompt transition into the detonation regime would be akin to the prompt scenario describe above - or after a number of "pulses" (the pulsational DDT scenario). These pulses would see one or more increases and subsequent drops in nuclear burning, accompanied by expansion and contraction of the WD, with CO being converted into IMEs, before the final ignition of a detonation which would then be visible as an SN.

Also included in this category are certain outcomes of systems resulting in a double detonation scenario (this mechanism is more relevant for the sub-Chandrasekhar category and is discussed in greater detail in that context. See Sec.~\ref{ssec:sub_ch}) as they continue to accrete to within $0.05~\text{M}_\odot$ of the Chandrasekhar mass \citep{NYL2019}, depending on the efficiency of angular momentum transport in the accreting WD.

As mechanisms acting at this terminal mass can be considered "classical", observed objects moving with velocities incompatible with terminal masses in the range $1.35~\text{M}_\odot>M_\text{d,f}>1.45~\text{M}_\odot$ will be most interesting.

\subsection{Sub-Chandrasekhar mass mechanisms} \label{ssec:sub_ch}

If thermonuclear SNe can be ignited well below the Chandrasekhar mass, then the initiation of nuclear burning will not be related to the WDs stability against gravitational collapse. The initial "spark" setting off the thermonuclear detonation of the CO core will therefore not be lit at the center of the WD. One mechanism capable of achieving this was first proposed in the 1980s by \cite{N1980P,N1982b}. This mechanism, now widely known as the double detonation (DDet) mechanism, posits that a helium layer, accumulated from a helium-rich companion star of some description, on top of the WD's CO core can act as a detonator for the thermonuclear disruption of the star. Under certain (currently relatively well but not completely understood) circumstances, ignition of nuclear burning in such a helium layer will lead to a detonation of the helium. The associated detonation shock will then trigger a secondary detonation of the carbon in the WD's core. 
Terminal accretor masses in this scenario have been argued to be as low as $0.75~\text{M}_\odot$ \citep{LA1995}.

Another mechanism falling into this category, while very similar to the DDet mechanism is the prompt double detonation (PDDet) or dynamically driven double degenerate double detonation (D6) mechanism. This mechanism relies on turbulent ignition of a thin $<0.05~M_\odot$ helium layer, accumulated dynamically from a companion (most likely a He-WD or hybrid HeCO WD), setting off a secondary detonation of the accretor's CO core \citep{PKT2013}. 

\subsection{Super-Chandrasekhar mass mechanisms}

In the DD scenario, super-Chandrasekhar mass explosions are expected to occur simply if the total mass of the system, containing two sub-Chandrasekhar mass WDs, is sufficiently high. In the SD scenario, however, theory requires super-Chandrasekhar mass detonations to rely on rotation. The effects of rotation on the stability of WDs against gravitational collapse has been of interest to the astrophysical community for some time. Specifically, \cite{YL2005} showed that fully differentially rotating WDs may avoid gravitational collapse at masses up to $2.2~\text{M}_\odot$. Rigid rotation is unable to prevent gravitation collapse at masses higher than $1.5~\text{M}_\odot$ \citep{HKSN2012}. In either case, an accreting WD, starting at masses sufficiently below the Chandrasekhar mass to allow for the WD to be sufficiently spun up by angular momentum accretion to avoid collapse at the Chandrasekhar mass can be expected to grow beyond it. 
After the system detaches, the fast rotation of the WD can then be slowed through mechanisms like tidal interaction, the r-mode instability \citep[see e.g.][]{YL2004b} or, conceivably, magnetic braking \citep{M1968}. As the rotation of the WD slows (which is expected to occur on timescales of up to Gyrs), it becomes progressively less stable against gravitational collapse, which is expected to occur once the WD has lost a certain amount of angular momentum. 
This scenario, widely known as the spin-up/spin-down (su/sd) mechanism was first proposed some time ago \citep[see e.g.][]{DVC2011}, with significant uncertainties remaining related to explosion physics and the spin-down timescale. 
However, it is unlikely that a runaway hot subdwarf is produced if the spin-down timescale is very long.

\begin{table*}
	\caption{Sources of utilized test cases}
	\label{table:sources}      
	\centering
	\begin{tabular}{ c c c c }
		\hline\hline
		Label & Citation & Method & Additional Notes \\    
		\hline                        
		N1982 & \cite{N1982b,N1982a} & 1D SEC & DDet mechanism proposed \\
		WW1994 & \cite{WW1994} & 1D SEC & - \\
		LA1995 & \cite{LA1995} & 2D HS & - \\
		YL2004 & \cite{YL2004b} & 1D SEC & Included rotational instabilities \\
		PCL2004 & \cite{PCL2004} & 2D HS & - \\
		GKO2005 & \cite{GKO2005} & 3D HS & Representative for the $M_\text{Ch}$-case \\
		YL2005 & \cite{YL2005} & 1D SEC & Super-$M_\text{Ch}$-case, differential rotation \\
		HSN2006 & \cite{HSN2006} & Observational & Observed Super-$M_\text{Ch}$, inferred mass \\
		FHR2007 & \cite{FHR2007} & 3D HS & - \\
		SRH2010 & \cite{SRH2010} & 1D HS & - \\
		LCW2010 & \cite{LCW2010} & 1D SEC & - \\
		WK2011 & \cite{WK2011} & 1D SEC+HS & - \\
		NYL2017 & \cite{NYL2017} & 1D SEC & included rotation+magnetic torques \\
		\hline
	\end{tabular}
\tablefoot{Sources for main test cases used in this study, including the methodology employed by each source: Stellar evolution calculation (SEC), hydrodynamical simulation (HS) or observations.}
\end{table*}

\section{Physical considerations} \label{sec:assumption}

The terminal orbital velocity of a donor star of a given mass $M_\text{d}$ and radius $R_\text{d}$ in a binary system undergoing a thermonuclear supernova can be calculated by the widely-used approximation provided by \cite{E1983}
\begin{equation} \label{eq:eggle}
a = \frac{R_\text{d,RL}}{0.49} \left\lbrace 0.6 + \left( \frac{M_{WD}}{M_\text{d}} \right)^{2/3} \ln \left[ 1 +\left( \frac{M_\text{d}}{M_\text{WD}} \right)^{1/3} \right] \right\rbrace 
\end{equation}
with the assumption that the donor star exactly fills its Roche lobe (i.e. the condition $R_\text{d}=R_\text{d,RL}$). The terminal orbital velocity then follow from the Keplerian equation
\begin{equation} \label{eq:kep}
v_\text{ej} = \sqrt{\frac{GM_\text{WD}^{2}}{(M_\text{WD}+M_\text{d})a}}
\end{equation}
where $M_\text{WD}=M_\text{acc}$ is the mass of the accretor, and $G$ is the gravitational constant. 
If the quantity $R_\text{d}(M_\text{d})$ can be expressed analytically, the terminal orbital velocity follows immediately from Eqs.~\ref{eq:eggle} and \ref{eq:kep}. However, as $R_\text{d}(M_\text{d})$ realistically depends on the structure of the star as well as its current mass loss rate due to RLOF, a consistent solution of both the stellar structure equations and the orbital evolution of the system, is required. 

Unlike previous studies \citep[e.g.][]{BSB2017,NYL2019}, which included the detailed evolution of the accreting companion, this study treats the accretor as a point mass. This, consciously adopted, simplification allows maintenance of a certain "agnosticism" towards the explosion mechanism and terminal state of the accretor.

The orbital evolution in binary systems is generally influenced by the effects of magnetic braking \citep{M1968}, tides \citep{H1981}, mass transfer and gravitational wave radiation \citep[GWR, see e.g.][]{LL1975}. 
Of these, magnetic braking is thought to be important mainly in solar-type stars and neglected in He rich stars like the ones under consideration here. As the systems under consideration here can be thought of as circularized and tidally locked subsequent to the most recent common envelope (CE) phase ("most recent" referring to the CE phase immediately preceding the formation of the He donor). Tides, as well as the possibility of non-conservative mass transfer, are neglected.

The orbital evolution of these systems therefore dominated by mass transfer and GWR. While angular momentum loss due to GWR acts to decrease orbital separation under all circumstances, the angular momentum transported during mass transfer may act to either increase or decrease orbital separation depending on the system's mass ratio, with $q=M_\text{d}/M_\text{acc}<1$ associated with increasing orbital separation and vice versa. 
In the systems under consideration, RLOF is induced either through an increase of the donor star radius or decrease of the orbital separation due to GWR. Once the system evolves into a semidetached state, further evolution of the orbit is driven by both GWR and angular momentum transfer.

\cite{E2011B} gives for GWR, neglecting mass transfer
\begin{equation}
	\frac{\dot{a}}{a}=-\frac{2}{\tau_\text{GR}}
\end{equation}
with $a$ the semi-major axis and the gravitational merger timescale
\begin{equation} \label{eq:adot-gr}
\tau_\text{GR} =\frac{5}{32} \frac{c^5 a^4}{G^3 (M_1+M_2) M_1 M_2}
\end{equation}
and for mass transfer, neglecting GWR
\begin{equation} \label{eq:adot-mt}
	\frac{\dot{a}}{a}=2 \frac{\dot{M}_1}{M_1+M_2} \frac{q^2-1}{q}.
\end{equation}
Note that Eq.~\ref{eq:adot-mt} implies that the evolution of the orbital separation is independent of the mass transfer rate if $q=1$. 
Comparing Eqs.~\ref{eq:adot-gr} and \ref{eq:adot-mt} yields, with the assumption $q<1$ and the demand that $\dot{a}/a > 0$ the condition
\begin{equation} \label{eq:adot-cond}
	\dot{M}_1 < \dot{M}_\text{crit} = - \frac{32}{5} \frac{G^3}{c^5 a^4} \frac{(M_1+M_2) M_1^2 M_2^2 }{M_1-M_2} .
\end{equation}
as derived by \cite{TY1979}.
Due to its inverse proportionality to the fourth power of $a$, Eq.~\ref{eq:adot-cond} is usually fulfilled by default in most systems containing an ordinary star, whose physical radius is on the order of one magnitude greater than that of a hydrogen depleted star of comparable mass.
With gravitational merger timescales being comparable to the mass transfer timescales ($ \tau_\text{MT} = \frac{\dot{M}_1}{M_1+M_2}$) in the systems under consideration here, the sign of the time derivative of the orbital separation will be determined by Eq.~\ref{eq:adot-cond}.

\section{Numerical methods and initial model parameters} \label{sec:methods}

The fundamental methodology of this study consists of the detailed calculation of the orbital evolution of close He-star+WD binaries. This is accomplished using the MESA framework \citep{MESA1,MESA2,MESA3,MESA4} in its release 11398.

MESA is publicly available and well established stellar and binary evolution code, capable of treating the evolution of single stars as well as that of the orbital parameters of binary systems. 
MESA offers a variety of prescriptions to calculate mass loss due to RLOF. For systems of this type, the mass loss prescription provided by \cite{R1988}, implemented as MESA-option \texttt{mdot\_scheme = 'Ritter'}, is considered most appropriate. This scheme relies on implicitly solving 
\begin{equation} \label{eq:ritter}
R_\text{d,RL} - R + H_P \cdot \ln \left( \frac{\dot{M}}{\dot{M}_0} \right) = 0~
\end{equation} 
where $H_P$ is the photospheric pressure scale height, $R$ the stellar radius as defined by the photosphere, and $R_\text{d,RL}$ the Roche lobe radius with $R_\text{RL}$ calculated according to Eq.~\ref{eq:eggle}.

This study is mostly concerned with the ejection velocity of donor stars ejected from systems with terminal accretor masses in the range $1.1~\text{M}_\odot \leq M_\text{t,acc} \leq 1.5~\text{M}_\odot$. In order to provide sufficient coverage of the grid, initial accretor masses were chosen in the range $0.5~\text{M}_\odot \leq M_\text{t,acc} \leq 1.2~\text{M}_\odot$ in steps of $0.05~\text{M}_\odot$.
Initial donor star models range in mass between $0.4~\text{M}_\odot$ and $1.0~\text{M}_\odot$ with solar metallicity.
As per Eq.~\ref{eq:eggle}, large stellar radii will lead to lower ejection velocities. Limiting this study to likely production mechanisms of hypervelocity stars originating in the Galactic disk, this condition excludes post-HeMS stars due to the requirement to exceed the Galactic escape velocity during ejection. Calculations are therefore terminated if the donor stars reaches the end of its core helium burning phase before the onset of RLOF.  

\subsection{Initial models} \label{sec:initmod}

\begin{figure} 
	\input{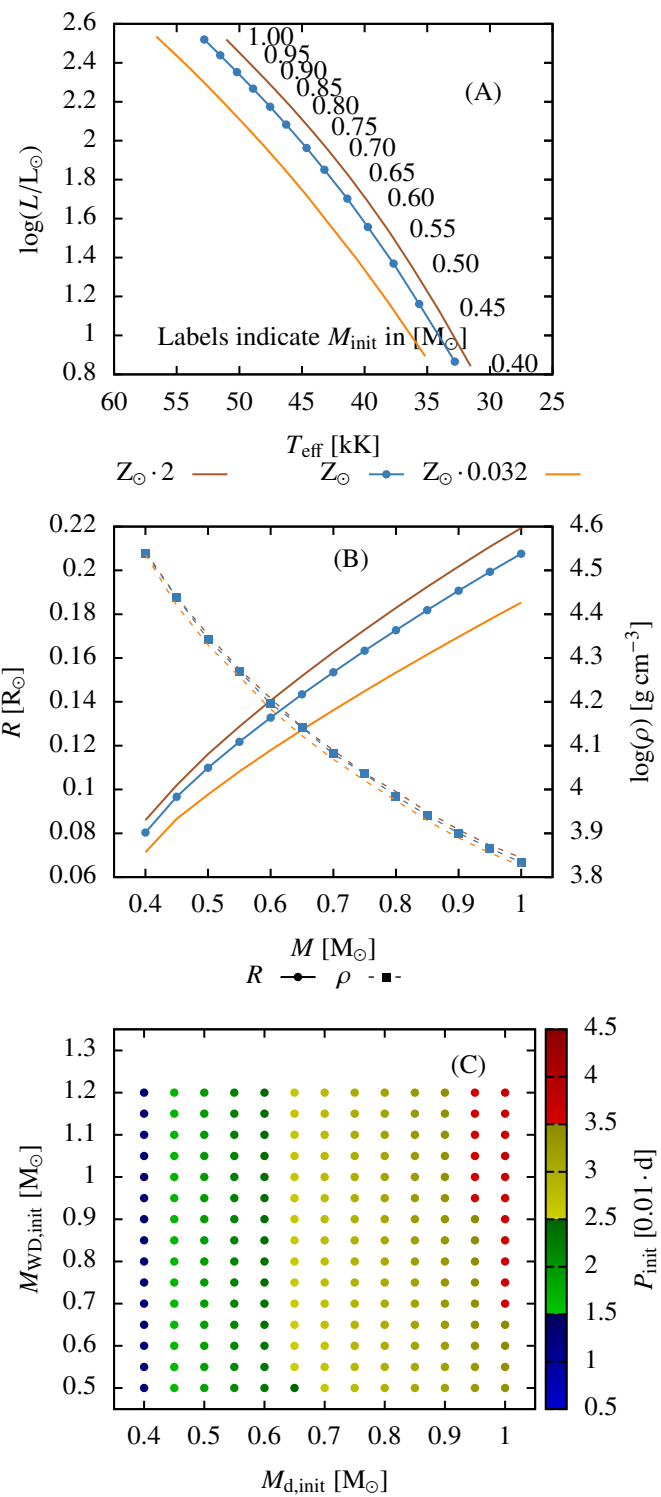}
	\caption{Relevant initial parameters of the utilized He-donor star models. Panel (A) indicates the position of each model in the HR-diagram, with labels corresponding to initial mass in units of $M_\odot$. Colors indicate initial metallicity in units of the solar metallicity $Z_\odot$ with solar metallicity used in this study. 
	Panel (B) indicates the initial radius $R_\text{init}$ in units of the solar radius $\text{R}_\odot$ and central density $\rho_\text{c}$ with colors corresponding to metallicity as in panel (A). Panel (C) shows initial periods, according to the constraints discussed in Sec.~\ref{sec:initmod} with respect to the initial masses of both components.}  \label{fig:initpars}
\end{figure}

The initial states of the employed donor models are summarized in Fig.~\ref{fig:initpars}.
These initial models are created by initializing a hydrogen-depleted pre-MS model using the MESA-supplied option \texttt{create\_pre\_main\_sequence\_model = .true.}, allowing it to settle on the HeMS, relaxing it further for a thermal timescale (He-rich models like this tend to contract by 1-5\% after reaching the HeMS). The donor model is then placed in a binary system with appropriate characteristics.
Initial orbital separations $a_\text{init}(\xi)$ were chosen such that Eq.~\ref{eq:eggle} satisfies $R_\text{d,RL}=\xi \cdot R_\text{d}$ with $\xi$ an arbitrary dimensionless parameter.
Depending on the individual component masses within the given ranges, a system of the present configuration is not expected to interact during the donor's HeMS lifetime at all for orbital radii $a_\text{init}(\xi > 1.01)$. The most viable choice is deemed to be $a_\text{init}(\xi = 1.005)$ with a secondary sample of $a_\text{init}(\xi = 1.01)$ (see Appendix~\ref{app:as}). This increase only results in a comparably small increase in the initial period of the system. However, as shown in \cite{NYL2019}, donor stars of initial mass $\gtrsim0.8~\text{M}_\odot$ tend to reach the end of core helium burning disproportionally quickly while those of lower mass $\lesssim0.5~\text{M}_\odot$ tend to evolve slowly enough to not have experienced significant increase in mass or metallicity before reaching GWR-induced RLOF. While this approach somewhat limits predictive power in intermediate masses, upper and lower limits should be adequately addressed. Initial periods resulting from this prescription are shown in Fig.~\ref{fig:initpars} (C).

\subsection{Stopping conditions}

Computations are terminated either when the donor star's core helium abundance decreases below $Y_\text{core} = 10^{-3}$ or if the donor star's mass drops below $0.18~\text{M}_\odot$. 
The first condition is motivated by the desire to focus on the production of hypervelocity hot subdwarfs. Further, the coincident expansion of their envelopes will result in increased mass transfer rates and, therefore, according to Eq.~\ref{eq:adot-cond}, an increase in orbital separation and, consequently, decreased ejection velocities. While further investigation of these hypothetical post-hot-subdwarf runaways is of some interest, it is beyond the scope of this paper.
The second condition is warranted due to the limitations of the equation of state (EOS) as currently implemented in MESA as donor star models in their proto-WD phase (i.e. $M<0.3\text{M}_\odot$) tend to cross into regions of the parameter space with insufficient coverage \citep[see][Appendix A]{MESA5}, resulting in numerical artifacts or unresolvable models. Specifically, these instabilities occur as sufficient amounts of unburnt helium from the outer layers of the star are removed, exposing the formerly burning and metal enriched core layers, which, at this point, will be cool and sparse enough lie outside the coverage of the EOS.
It is found that this problem is largely avoided by stopping the simulation at $0.18~\text{M}_\odot$, as most models with both more massive and significantly metal enriched cores will have left the HeMS (and, consequently, been removed by the first stopping condition) at this point with remaining models possessing either less massive or sufficiently pristine cores.

\section{Ejection velocity spectra} \label{sec:results}

The principal aim of this study is to provide ejection velocity spectra for hypervelocity runaways produced in the He-star+WD channel for thermonuclear SNe. An ejection velocity spectrum, for the purposes of this paper, is composed of the expected ejection velocities of a runaway, depending on the terminal mass of the runaway, for a single terminal WD mass. 
In order to remain unbiased towards the plethora of proposed explosion mechanisms, as discussed in Sec.~\ref{sec:expl}, which would impose systematic constraints on the derived ejection velocity spectra, a range if terminal WD masses are taken into consideration. 

\subsection{Partial spectra}
\begin{figure} 
	\input{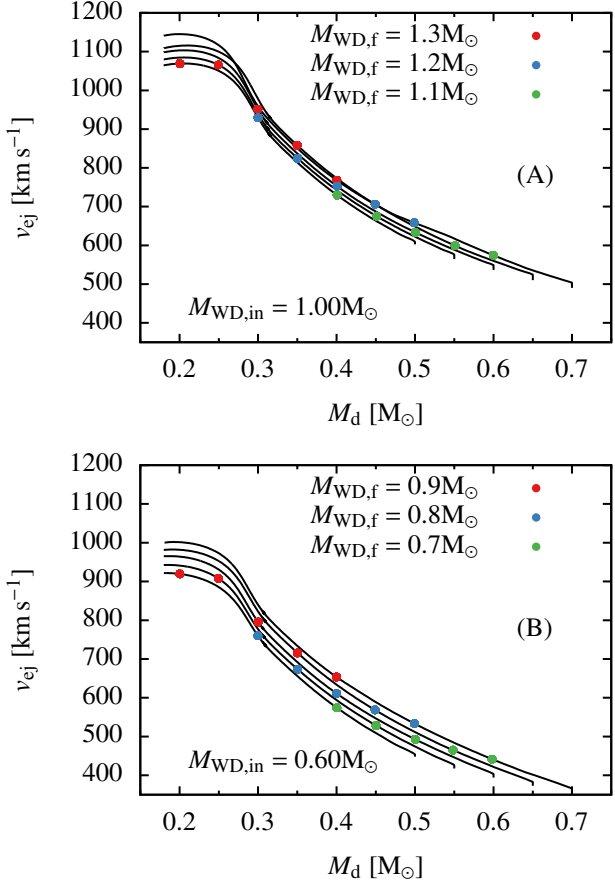}
	\caption{Representative model sequences detailing the evolution of individual systems in the $M_\text{d}$-$v_\text{ej}$ space. $M_\text{WD,in}$ is the initial accretor mass for each of the model sequences with $M_\text{WD,in} = 1.00~\text{M}_\odot$ in panel (A) and $M_\text{WD,in} = 0.60~\text{M}_\odot$ in panel (B). Each dot represents a single model with terminal accretor mass $M_\text{WD,f}$. Velocities given with respect to the center of mass of the progenitor binary} \label{fig:v_bmass_singles}
\end{figure}

 Contribution of individual binary model sequences to the ejection velocity spectra for any particular terminal WD mass ($M_\text{WD,f}$) are shown in Fig.~\ref{fig:v_bmass_singles}. The entire spectrum for any particular choice of $M_\text{WD,f}$, as derived in this study is then composed of all binary system states with the same value of $M_\text{WD,f}$. This means that, rather than being determined by the evolution of the orbital velocity of an individual binary model sequence, the ejection velocity spectrum is composed of the systems' in the $v_\text{orb}$-$M_\text{d}$ parameter space where the mass of the accretor is equal to the requested value of $M_\text{WD,f}$ across multiple model sequences.
In the partial spectra seen in Fig.~\ref{fig:v_bmass_singles}, a clear correlation between orbital velocities and the terminal mass of the accretor, independent of both initial masses is apparent. Individual tracks exhibit a maximum of the orbital velocity, located in the range $0.2~\text{M}_\odot < M_\text{d} < 0.25~\text{M}_\odot$. As the full ejection velocity spectra are essentially determined by the shapes of numerous individual tracks, this feature is also expected in the complete spectra. 

\subsection{The ejection velocity maximum}
\begin{figure}
	\input{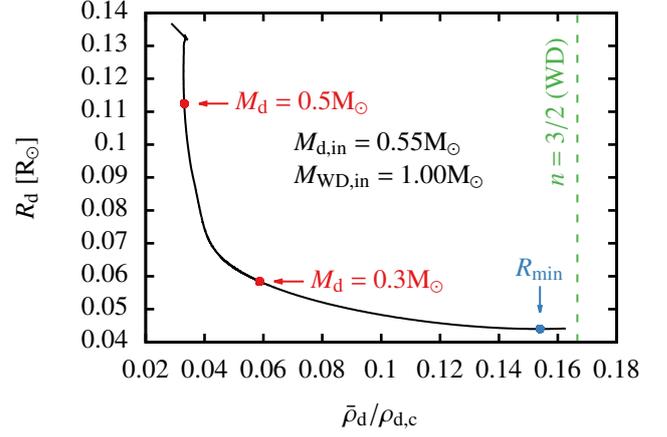}
	\caption{Radius of the donor star ($R_\text{d}$) in relation to the quotient ($\bar{\rho}_\text{d}/\rho_\text{d,c}$) for the same system as in Fig.~\ref{fig:compv}. The minimum radius $R_\text{min}$ is reached at a mass of $0.20~\text{M}_\odot$ in this model sequence.} \label{fig:Rcompact}
\end{figure}
As argued previously, the maximum in ejection velocities exhibited by individual model sequences is expected to translate to the full spectra. This indicates that for each assumed value of $M_\text{WD,f}$ there exists a maximum ejection velocity $v_\text{ej,max}(M_\text{WD,f})$. Consequently, any observed hypervelocity runaway with an inferred ejection velocity higher than $v_\text{ej,max}(M_\text{WD,f})$ for an assumed $M_\text{WD,f}$ must necessarily have been ejected either from a system with a higher $M_\text{WD,f}$ or by a different mechanism altogether. Any such observation is especially auspicious in the case of $M_\text{WD,f} \sim 1.4~\text{M}_\odot$. 
\begin{figure}
	\input{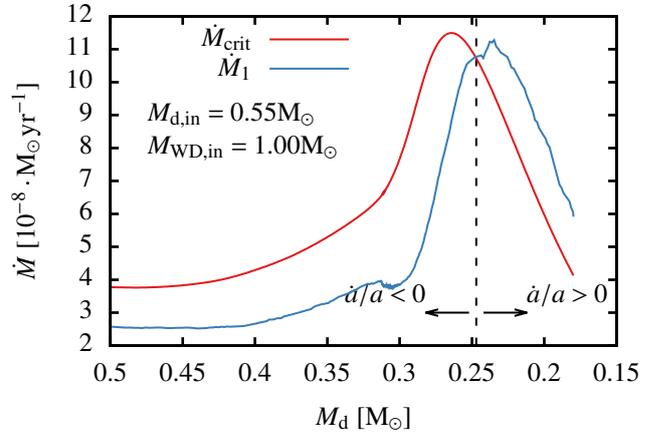}
	\caption{Comparison between the critical mass transfer rate $\dot{M}_\text{crit}$ for increasing $a$ as defined in Eq.~\ref{eq:adot-cond} and the actual mass transfer rate $\dot{M}_1$ for a representative system with respect to donor mass $M_\text{d}$. Initial masses $M_\text{d,in}$ and $M_\text{WD,in}$ are as indicated.} \label{fig:compv}
\end{figure}
The occurrence of this maximum is driven by the widening of the binary as the donor star becomes more degenerate once its mass drops below the value needed for sustained helium burning. This is the  $0.3~\text{M}_\odot$-limit mentioned above.
As the donor star loses sufficient mass to drop below this limit, the lack of energy generation by nuclear fusion leads to contraction on the thermal timescale concurrent with a reconfiguration of the star's structure to become more degenerate. This reconfiguration is conveniently illustrated by the evolution of the quotient $\bar{\rho}_\text{d}/\rho_\text{d,c}$ (i.e. the donor star's average density divided by its central density). This quotient is a well known quantity in polytropic stellar models with 
\begin{equation}
	\frac{\bar{\rho}}{\rho_\text{c}}=\left( -\frac{3}{z} \frac{dw}{dz} \right)_{z=z_n}
\end{equation}
Where $z_n$ are the solutions to the Lane-Emden equation
\begin{equation}
	\frac{1}{z^2}\frac{d}{dz}\left(z^2 \frac{dw}{dz} \right) + w^n = 0
\end{equation}
 and $n$ the polytropic index. Low mass WDs are well approximated by polytropic equations of state with $n=3/2$ \citep[see e.g.][]{STELLAR_STRUCTURE_AND_EVOLUTION}.
In the systems under discussion here, as the donor star loses mass, the quotient $\bar{\rho}_\text{d}/\rho_\text{d,c}$ is thus expected to evolve towards values compatible with a polytropic index of $n=3/2$. An example of this is shown in Fig.~\ref{fig:Rcompact}. It is noteworthy that, as the donor star approaches $\left(\bar{\rho}_\text{d}/\rho_\text{d,c}\right)_{z_n=z_{3/2}}$, its radius passes a minimum. This inverse correlation (compared to main sequence stars) of mass and radius is a well known property of WDs.
\begin{figure} 
	\input{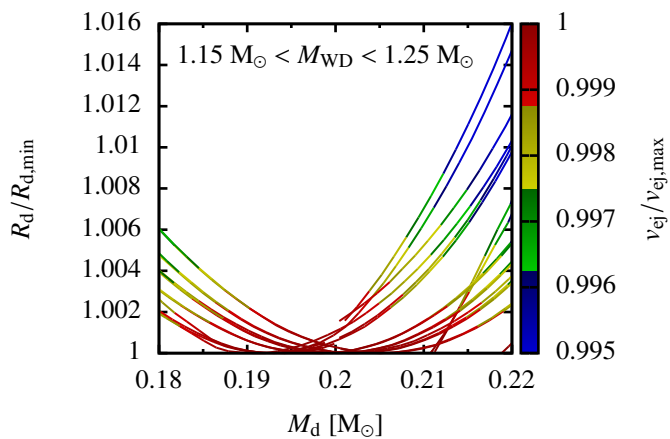}
	\caption{Mass-radius relationship of donor stars in the vicinity of the ejection velocity maximum. The mass-dependent donor star radius ($R_\text{d}$) is normalized to the minimum radius of that particular model sequence ($R_\text{d,min}$), the mass dependent ejection velocity, indicated by the color bar, ($v_\text{ej}$) is analogously normalized to the maximum ejection velocity of the same model sequence. For clarity, only tracks in the vicinity of an accretor mass in the range $1.15~\text{M}_\odot < M_\text{WD} < 1.25~\text{M}_\odot$ are shown.} \label{fig:vRL}
\end{figure}
As the donor star approaches a radius minimum, the mass transfer rate, as shown in Fig.~\ref{fig:compv}, increases. As shown, this increase in mass transfer is sufficient to exceed the critical mass transfer rate defined in Eq.~\ref{eq:adot-cond}, in turn leading to an increase of the orbital separation and, hence, a decrease of the orbital and ejection velocity. 
In the present sample, the radius minimum is generally attained in the range $0.19~\text{M}_\odot < M_d < 0.25~\text{M}_\odot$ (Fig.~\ref{fig:vRL}). Objects in this range would, as indicated by Fig~\ref{fig:Rcompact}, be characterized as proto-WDs that would, following a period on further contraction, form a population of low-mass, high velocity runaway WDs. However, as these objects are likely to only properly settle on the WD cooling sequence a considerable time after ejection and, corresponding to their high ejection velocity, a significant distance from their point of origin, observation of such an object as a high velocity extremely low mass (ELM) WD is deemed unlikely, though not impossible. This scenario is similar to the one proposed by \cite{JWP2009} but has to be tempered with the notion of a large fraction if not all of the currently observed ELM WDs being part of a binary \citep{BKK2020}.

\subsection{Bifurcations} \label{sec:bifurcations}
\begin{figure} 
	\input{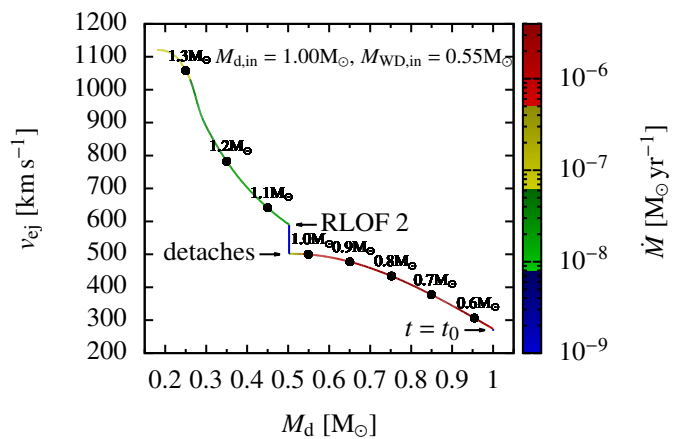}
	\caption{Model sequence in $M_\text{d}$-$v$-space of a single binary system with initial masses $M_\text{d,in}$ and $M_\text{WD,in}$, i.e. $q>1$. Color indicates the current mass transfer rate $\dot{M}$. Labels along the graph indicate the current accretor mass. $t=t_0$ is indicated as well as the points where the system detaches and undergoes a second RLOF. Velocity given with respect to the center of mass of the progenitor binary} \label{fig:v_bmass_q_single}
\end{figure}
The orbital velocity evolution of a system with initial mass ratio $q>1$ are shown in Fig.~\ref{fig:v_bmass_q_single}. As can be seen, the system undergoes two distinct phases of mass transfer: A "fast" phase of high mass transfer rates, followed by a "slow" phase of low mass transfer rates. In stars only undergoing core burning, the mass transfer timescale is generally comparable to the donor star's nuclear timescale (i.e. $\tau_\text{nuc}\sim\tau_\text{RLOF}$), however, in systems with $q>1$, angular momentum transfer due to RLOF additionally acts to decrease the system's orbital separation, leading to enhanced mass transfer. 
However, as the donor star is initially the more massive companion in these systems, it will initially orbit with a lower orbital velocity. The donor star will lose mass to the accretor until $q=1$ is reached, at which point angular momentum transfer will act to increase the orbital separation (see Eq.~\ref{eq:adot-mt}) until the system detaches. Prior to the point of detachment, the orbital velocity of the donor star will generally be lower than in systems with $q<1$ at the same donor star mass. Subsequent to detachment, the components will then evolve in isolation until angular momentum loss due to GWR has decreased the orbital separation sufficiently to initiate a second RLOF. The lower orbital velocities prior to detachment will lead to the presence of a secondary branch in the full ejection velocity spectra. However, due to the limited nature of the initial parameter space in this study, this secondary branch is only resolved in the spectra corresponding to the lowest terminal WD masses.

\subsection{Complete spectra} \label{sec:fullspec}
\begin{figure*} 
	\includegraphics{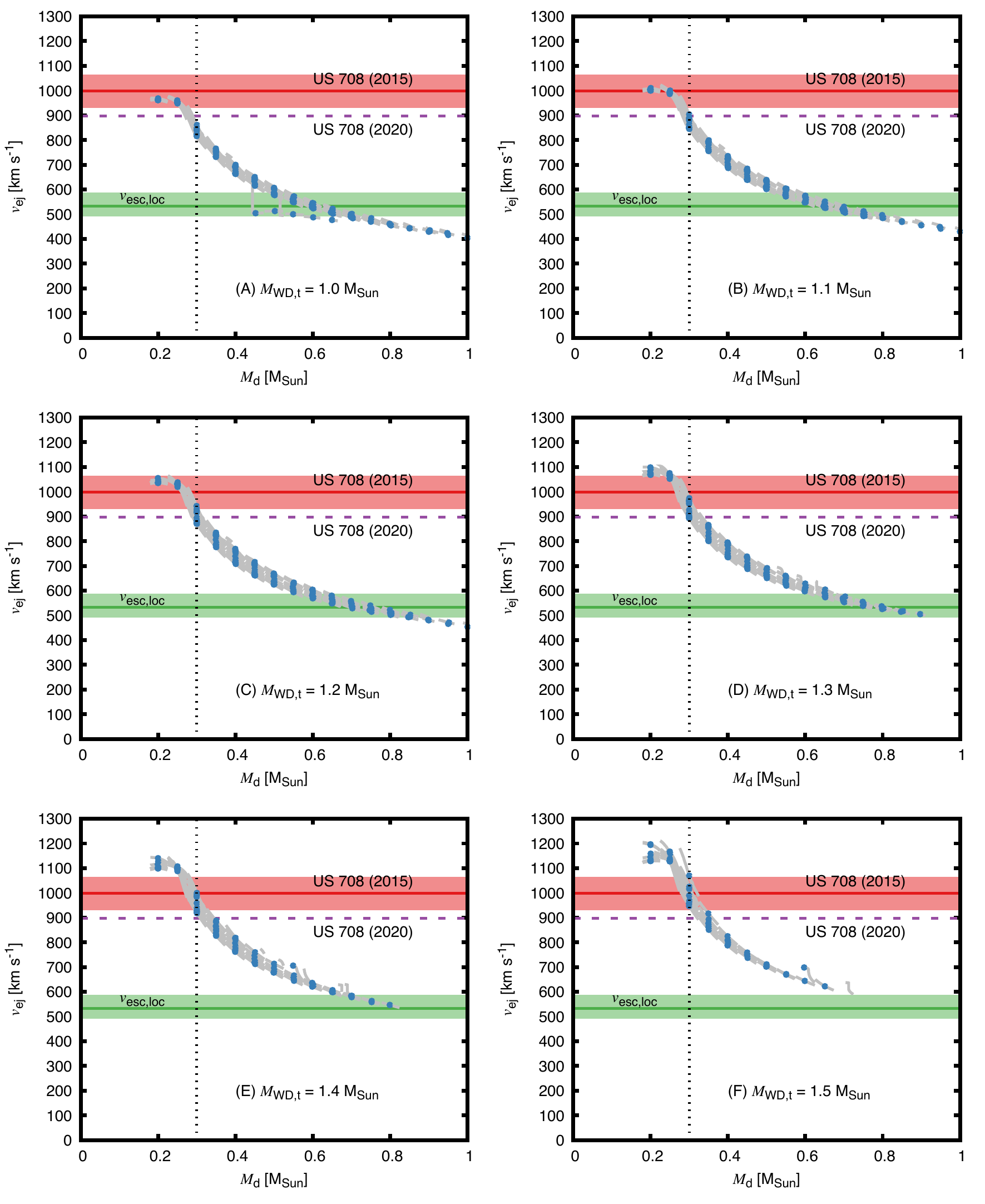}
	\caption{Ejection velocity spectra for indicated terminal accretor masses ($M_\text{WD,t}$). $M_\text{d}$ and $v_\text{ej}$ are the terminal mass and expected ejection velocity with respect to the center of mass of the progenitor binary of the remnant runaway hot subdwarf. The blue dots indicate the exact state of the system when the accretor reaches the indicated $M_\text{WD,t}$. Lines indicate the evolution of the system in an envelope from $M_\text{WD,t}-0.025\cdot\text{M}_\odot < M_\text{WD,t} < M_\text{WD,t}+0.025\cdot\text{M}_\odot$. The dotted purple line indicates $M_\text{rem} = 0.3~\text{M}_\odot$, below which core helium burning ceases. The inferred ejection velocity of US 708 according to \cite{GFZ2015} and current proper motions provided by Gaia-DR2 \citep{GAIA-DR2}, represented by the solid red and dashed purple lines respectively, and local Galactic escape velocity according to \cite{PSB2014}, including error bars, are given for orientation.} \label{fig:v_bmass}
\end{figure*}
Full ejection velocity spectra in the range $1.0~\text{M}_\odot < M_\text{WD,f} < 1.5~\text{M}_\odot$ are shown in Fig.~\ref{fig:v_bmass}. Note that, as would naively be expected, the maximum ejection velocity is correlated with $M_\text{WD,f}$ and the presence of an ejection velocity maximum in the range $0.19~\text{M}_\odot < M_d < 0.25~\text{M}_\odot$ for all values of $M_\text{WD,f}$. In panel (A) with $M_\text{WD,f} = 1.0~\text{M}_\odot$, a bifurcation of the ejection velocity spectrum, as described above, is visible. 
It should further be noted that, with increasing values of $M_\text{WD,f}$ the maximum donor star mass shown in each panel decreases. This is only partially a consequence of the choice of initial parameter space, as systems that do not interact during the helium main sequence of the donor star are excluded from this plot. Donor stars of high initial mass tend to evolve quickly enough to avoid RLOF during the helium main sequence at the initial orbital separations chosen in this study and are therefore removed from the sample. For these systems to produce an SN during the core helium burning stage, the donor would have to full its Roche lobe entirely directly subsequent to the most recent CE phase.

Systems in the entire considered range of $M_\text{WD,f}$ are capable of producing hypervelocity stars, assuming ejection occurs in the Solar neighborhood, higher mass runaways ($\gtrsim 0.5~\text{M}_\odot$) are less likely to become unbound from the Galaxy depending on both the terminal WD mass, the local Galactic escape velocity and ejection direction. As such, the local Galactic escape velocity should be compared with the ejected companion's space velocity immediately following the SN event.
Further, as seen in Fig.~\ref{fig:v_bmass} (E), local Chandrasekhar mass explosions can be expected to always produce a hypervelocity hot subdwarf.
Ejection velocities higher than $1000~\text{km\,s}^{-1}$ can be expected for terminal WD masses $\gtrsim1.1~\text{M}_\odot$.
The spread in the ejection velocity spectra (i.e. the presence of multiple ejection velocities for a single value of $M_\text{d}$) is a consequence of the degeneracy of multiple initial systems for a set of terminal WD masses and terminal donor masses. The spread is then a consequence of each donor star in the degenerate set having lost a different amount of mass to the accretor during a mass transfer episode of different length, leading to a slightly different structure and chemical composition in each case. However, the spread is small enough that a dependence of the ejection velocity on the runaway mass is still clearly indicated.
The inferred ejection velocity of the runaway hot subdwarf US 708 according to \cite{GFZ2015} is $998~\text{km\,s}^{-1}$. This ejection velocity can be reached by any explosion involving an accretor mass $M_\text{WD,f}=1.1~\text{M}_\odot$. However, importantly, only if $M_\text{d,f}>0.3~\text{M}_\odot$ in systems with $M_\text{WD,f}\geq1.4~\text{M}_\odot$. 

With an inferred ejection velocity calculated from the proper motions provided in Gaia-DR2, \citep{GAIA-DR2} of $897~\text{km\,s}^{-1}$, the terminal accretor mass could be as low as $M_\text{WD,f}=0.85~\text{M}_\odot$. In this case, the minimum terminal accretor mass for $M_\text{d,f}>0.3~\text{M}_\odot$ is, notably, $M_\text{WD,f}=1.1~\text{M}_\odot$. The case of US 708 will be discussed in greater detail in Sec.~\ref{sec:US708}.

As seen in Fig.~\ref{fig:v_bmass} (F), $1.5~\text{M}_\odot$ explosions are capable of propelling a runaway to velocities up to $1200~\text{km\,s}^{-1}$, which is about $100~\text{km\,s}^{-1}$ slower than the D6-2 object as found by \cite{SBG2018}. However, it should be mentioned that the reliability of this particular measurement is being debated in literature \citep{S2018}.

\subsection{Runaway velocities as a probe of the pre-explosion progenitor state}
\begin{figure}
	\input{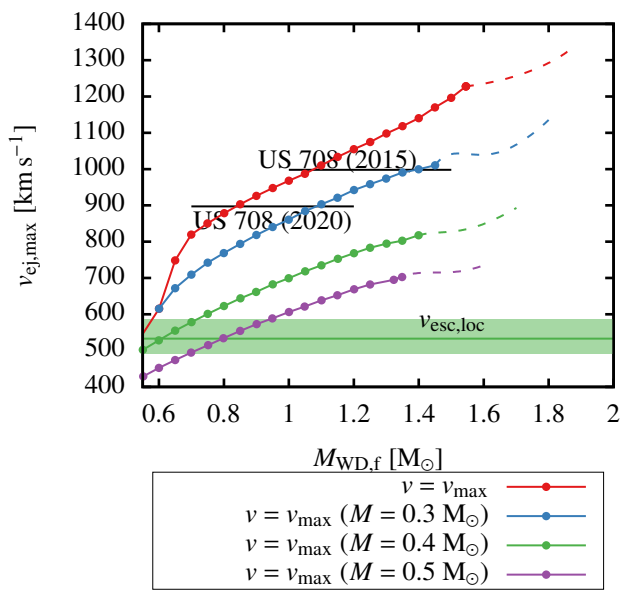}
	\caption{Maxima of all ejection velocity spectra as given in Fig.~\ref{fig:v_bmass} with respect to terminal accretor mass ($M_\text{WD,f}$), where $v_\text{max}$ indicates the absolute maximum and $v_\text{max}(M_\text{d})$ the maximum, with respect to the center of mass of the progenitor binary, for a given mass of the ejected companion ($M_\text{d}$). Black lines indicate inferred ejection velocities as calculated by \cite{GFZ2015} and using proper motions as provided by GAIA DR2 \citep{GAIA-DR2}. The dashed lines indicate the expected continuation of the maxima towards areas of the parameter space insufficiently covered by the grid.}  \label{fig:hsearch}
\end{figure}
As seen in Sec.~\ref{sec:fullspec}, given a constant value $M_\text{WD,f}$, ejection velocities are a strong function of $M_\text{d,f}$. Fig.~\ref{fig:hsearch} shows the maximum ejection velocities obtained in this sample with respect to the terminal WD mass (both the total maximum and the maxima for a given minimum mass of the runaway). As the depicted values are the theoretical maxima of ejection velocities, they can be used to obtain constraints on the parameters of the progenitor binary at the time of explosion. Specifically, since higher ejection velocities require higher terminal WD masses, but higher terminal donor star masses inhibit them, knowledge of the ejection velocity provides constraints on the parameter space of both the donor and the accretor mass. US 708 is given as an example in this plot, with a terminal WD mass in the range $1.07~\text{M}_\odot<M_\text{WD,f}<1.39~\text{M}_\odot$ with corresponding values of $0.18~\text{M}_\odot<M_\text{WD,f}<0.31~\text{M}_\odot$ for the terminal donor star mass based on proper motions obtained by \cite{GFZ2015}. 
Based on proper motions obtained from Gaia-DR2 \citep{GAIA-DR2}, these values amend to $0.85~\text{M}_\odot<M_\text{WD,f}<1.39~\text{M}_\odot$ with corresponding values of $0.18~\text{M}_\odot<M_\text{WD,f}<0.35~\text{M}_\odot$. In both cases assuming sub-Chandrasekhar or Chandrasekhar-mass explosions.
Note that, as calculated by \cite{BWB2019}, the terminal donor mass may be as much as twice that of the current mass of the eventually observed runaway.

\subsection{Rotation}

\begin{figure*} 
	\includegraphics{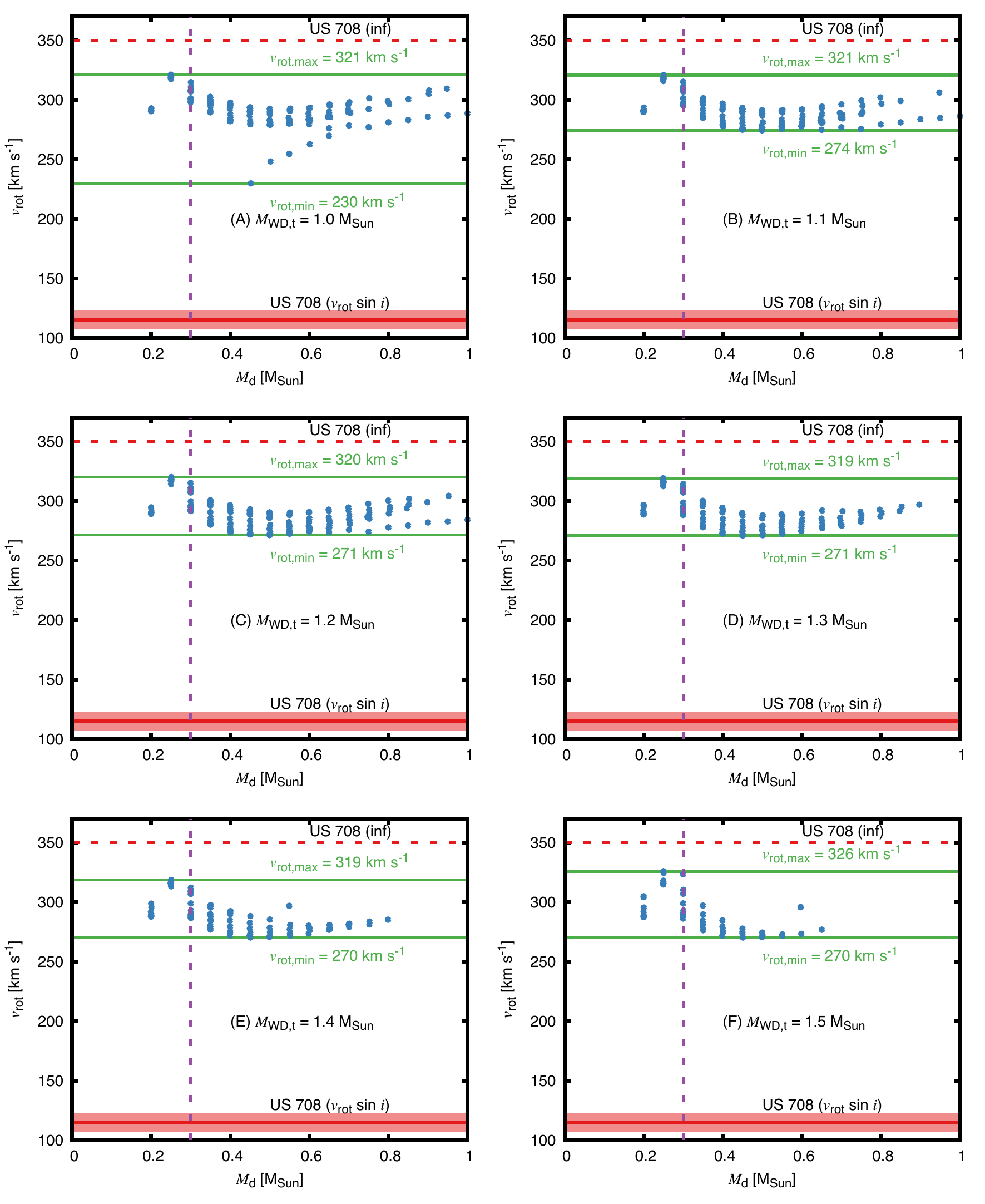}
	\caption{Same as Fig.~\ref{fig:v_bmass}, but showing the ejected companion's surface rotational velocity at the time of ejection with the observed current $v_\text{rot} \sin i$, including error bars, given for comparison, wit the red dashed line indicates the inferred surface rotational velocity at ejection according to \cite{GFZ2015}.
	Green lines indicate the minimum and maximum velocity found in the sample as labeled.} \label{fig:rot_bmass}
\end{figure*}

While the question of the rotational velocity of the ejected HVS is not a primary subject in this study, the fact of its potential accessibility to observation merits a brief discussion.
The terminal surface rotational velocities of the ejected component are shown in Fig.~\ref{fig:rot_bmass}. Excepting bifurcations, as discussed in Sec.~\ref{sec:bifurcations}, rotational velocities are expected to be uniformly higher than $v_\text{f,rot} = 270~\text{km\,s}^{-1}$ and lower than $v_\text{f,rot} = 326~\text{km\,s}^{-1}$, independent of the terminal accretor mass. With respect to the critical rotational velocity $ v_\text{rot,crit}$, this corresponds to a range between $0.29 \cdot v_\text{rot,crit}$ and $0.33 \cdot v_\text{rot,crit}$. The reason for the uniformity of rotational velocities lies in the underlying assumption of tidal locking, as donors of equal mass but increasing accretor masses will find themselves in correspondingly wider systems. This uniformity was also noted by \cite{GFZ2015}, with the upper limit presented in this study in reasonably good agreement with the inferred, model dependent, initial rotational velocity of $v_\text{f,rot} = \sim350~\text{km\,s}^{-1}$ derived by \cite{GFZ2015}. The assumption of tidal locking necessarily leads to a direct correlation of rotational and ejection velocity. The differing rotational velocity predicted at the lower limit (like the spread seen in the ejection velocity spectra) indicates that the donor star's structure and evolutionary history does have a noticeable effect on the final state of the system.
Remarkably, the spread of rotational velocities is comparable to that of ejection velocities, as shown in Fig.~\ref{fig:v_bmass}.
As can further be seen, the predicted terminal rotational velocities are larger by at least a factor of 2.3 than the observed current $v_\text{rot} \sin i = 115 \pm 8 \text{km\,s}^{-1}$ of US 708. Here it should be borne in mind that the radius of the ejected donor star is unlikely to correspond to the rotational velocity post-ejection. This was also noted by \cite{GFZ2015}. The decrease in velocity can be explained by the star's post-ejection evolution on the extreme horizontal branch under conservation of angular momentum without invocation of SN interaction. However, it should also be emphasized that the post-ejection evolution is likely strongly dependent on the question of ongoing nuclear processes inside the star, with stars below the threshold for helium burning reacting differently than stars above this threshold. Therefore, the thermal response to ejecta impact may still be important, as indicated by the results presented by \cite{BWB2019}.

\section{The case of US 708} \label{sec:US708}
As mentioned in Sec.~\ref{sec:introduction}, the hypervelocity runaway US 708 (HV2, SDSS J093320.86+441705.4, Gaia DR2 815106177700219392) is classified as a helium-rich hot subdwarf. Stars of this type are found at the blue end of the horizontal branch and are thought, most importantly for the purposes of this study, though not exclusively, to products of close binary evolution. Extensive discussion of the properties of these objects is beyond the scope of this paper, but for thorough reviews, the reader is directed to \cite{H2009} and \cite{H2016}.

In \cite{GFZ2015}, a current mass of $0.3~\text{M}_\odot$ for US 708, (due to the unavailability of a reliable mass measurement) was adopted, yielding a most likely terminal WD mass of $1.3~\text{M}_\odot$. With the predictions presented in Sec.~\ref{sec:results}, this assumption can be checked for consistency. This is done via the following prescription:
\begin{figure*}
	\input{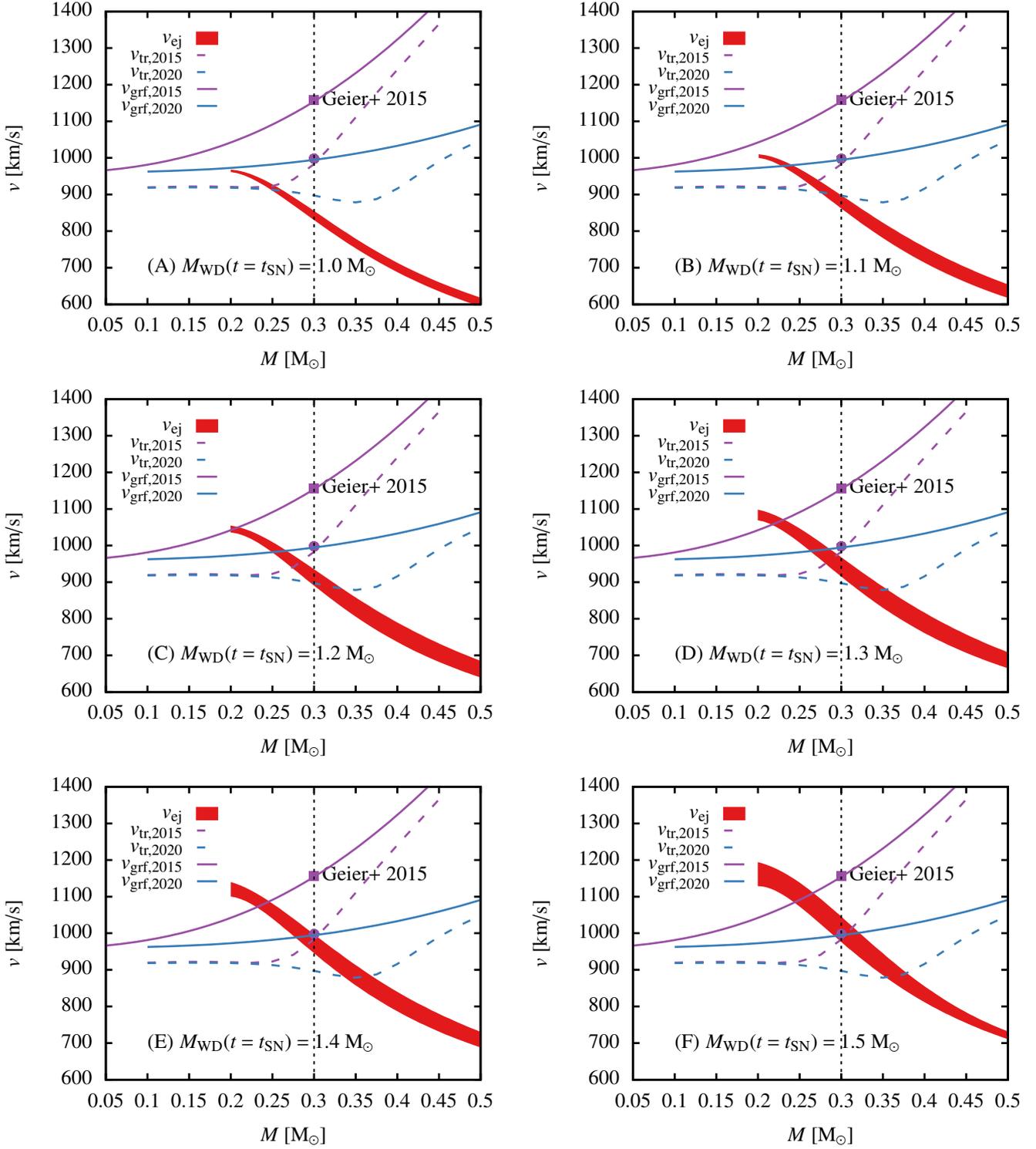}
	\caption{Inferred current Galactic rest frame velocity, $v_\text{grf}(t=t_0)$, and ejection velocity, $v_\text{tr}(t=t_\text{SN})$, of US 708, as dictated by observed radial velocity and proper motion, correlated with the ejection velocity spectra, $v_\text{ej}(t=t_\text{SN})$, with a WD explosion mass ($M_\text{WD}(t=t_\text{SN})$) as labeled. Observed space and ejection velocities are base either on \cite{GFZ2015} (denoted by subscript ''2015'') or on proper motions according to Gaia DR2 \citep[subscript ''2020'',][]{GAIA-DR2} Spectra are represented by the envelopes of their depictions in Fig.~\ref{fig:v_bmass} for legibility. The current space and inferred ejection velocities, as determined by \cite{GFZ2015} are depicted as purple dots.} \label{fig:US708ej}
\end{figure*}
\begin{enumerate}
	\item The current mass of US 708 is allowed to be a free parameter. This will, as distance was determined spectroscopically by \cite{GFZ2015}, impact the determination of the current space velocity, now relying solely on proper motion and radial velocity.
	\item \label{lab:1} Using kinematic analysis (See App.~\ref{app:traj}) and assuming ejection in the Galactic disc, a mass-dependent inferred ejection velocity, taking into account local Galactic rotation, is obtained.
	\item The resulting ejection velocities are compared with the ejection velocity spectrum for a given terminal WD mass.
\end{enumerate}
Regarding the second item, it should be mentioned that assuming ejection in the Galactic disc is not necessarily warranted, as the star may originate outside the disc, e.g. in a globular cluster, correspondingly affecting inferred ejection velocities.
Assuming similar structure, mass ($M$) and luminosity ($L$) of two stars are approximately correlated through the homology relation \citep[see e.g.][]{STELLAR_STRUCTURE_AND_EVOLUTION}
\begin{equation} \label{eq:hom-rel}
	\frac{L_1}{L_2} = \left( \frac{M_1}{M_2} \right)^3 \left( \frac{\mu_1}{\mu_2} \right)^4
\end{equation}
with $\mu$ the mean atomic weight, which is assumed to remain unchanged in this instance. Using the luminosity as calculated in Eq.~\ref{eq:hom-rel}, the distance can then be adjusted by assuming a constant apparent magnitude. 

As observational characterization of US 708 has advanced somewhat since 2015, with new proper motion data published in the Gaia data release 2 \citep{GAIA-DR2}, but reliable parallax distances and updated radial velocities still unavailable, the following kinematic analysis is performed using two distinct sets of proper motion parameters. The first set relies on the data obtained by ground-based observations published in \cite{GFZ2015} (labeled ''2015'' throughout the paper), the other on the more recent values obtained by the Gaia instrument and published in \cite{GAIA-DR2} (labeled ''2020'' throughout the paper). Numerical values are given in Tab.~\ref{tab:propmot}.
\begin{table} 
	\caption{Proper motions utilized in kinematic analysis} \label{tab:propmot}
	\begin{tabular}{ccc} 
		\hline 
		\hline
		 Data set & $\mu_\alpha \cos(\delta)$ & $\mu_\delta$    \\  
		  & [$\text{mas\,yr}^{-1}$] & [$\text{mas\,yr}^{-1}$]   \\
		\hline
		2015& $-8.0 \pm 1.8$  & $9.1 \pm 1.6$  \\ 
		2020& $-5.363 \pm 0.391$  & $1.285 \pm 0.382	$  \\
		\hline 
	\end{tabular} 
\end{table}

In the absence of accessible newer data, the values of visual magnitude ($m_g = 18.668 \pm 0.008\,\text{mag}$) and heliocentric radial velocity ($v_\text{helio} = 917 \pm 7\,\text{km\,s}^{-1}$) are adopted unchanged from \cite{GFZ2015}. As the determination of proper motion and radial velocity does not rely on an adopted mass or luminosity, they are kept constant in the calculation of the mass-dependent space velocity.
The reader is cautioned to note that the following kinematic analysis is intended to represent an idealized model, neglecting error propagation from observations and the Galactic potential. It is likely that inclusion of errors will detract from the unambiguity of the drawn conclusions.

\subsection{The 2015 data set}
Following the mass-dependent trajectories back to the Galactic plane-crossing yields an upper limit for the mass of US 708 of $0.45~\text{M}_\odot$ as any assumed higher mass would result in the trajectory avoiding intersection with the plane altogether.
A comparison of inferred current space and ejection velocities with predicted ejection velocity spectra is shown in Fig.~\ref{fig:US708ej}. 
Excluding additional momentum being imparted on the ejected runaway through interaction with the supernova ejecta, an assumed mass of $0.3~\text{M}_\odot$ is inconsistent with an terminal WD mass of $1.3~\text{M}_\odot$ (Fig.~\ref{fig:US708ej} D). \cite{BWB2019} suggest that ejecta interaction would impart an additional kick of $\sim 180~\text{km\,s}^{-1}$ on a runaway of $0.344~\text{M}_\odot$. Since this kick would be imparted perpendicular to the donor's orbital motion, the ejection velocity would be increased by $\sim 17~\text{km\,s}^{-1}$, still insufficient to allow for consistency. It can therefore be concluded that the mass of US 708 would need to be in the range $0.27~\text{M}_\odot < M_\text{US708} < 0.29~\text{M}_\odot$ assuming a terminal WD mass of  $1.3~\text{M}_\odot$.
Assuming a terminal WD mass of $1.4~\text{M}_\odot$, i.e. close to the Chandrasekhar mass, yields a likely range of $0.28~\text{M}_\odot < M_\text{US708} < 0.3~\text{M}_\odot$, consistent with the mass adopted by \cite{GFZ2015}. 
Assuming a super-Chandrasekhar mass explosion with a terminal WD mass of $1.5~\text{M}_\odot$ would indicate a mass range of $0.29~\text{M}_\odot < M_\text{US708} < 0.32~\text{M}_\odot$, also consistent with the mass adopted by \cite{GFZ2015}.
It can be concluded that, if US 708 was ejected in a sub-Chandrasekhar mass SN, then both its terminal and current mass should be smaller than $0.3~\text{M}_\odot$ and it should not currently burn helium. If, on the other hand, US 708 was ejected in a Chandrasekhar or super-Chandrasekhar mass SN, the question of its current state depends on the amount of material stripped during the SN event, it being highly likely that its terminal mass was greater than $0.3~\text{M}_\odot$. 
It can further be concluded that the observational properties of US 708 in the 2015 data set, including its inferred mass, is most consistent with an origin in a Chandrasekhar or super-Chandrasekhar mass SN. This is in agreement with the conclusions of \cite{GFZ2015}.

\subsection{The 2020 data set}

The Gaia-based proper motions imply a significantly lower current space velocity of around $994~\text{km\,s}^{-1}$, leading to an inferred ejection velocity (again assuming ejection in the Galactic disc) of around $897~\text{km\,s}^{-1}$ at a location both closer to Earth and the Galactic center than in the 2015 data set (see Fig.~\ref{fig:US708ejloc}). 
In this case, the maximum terminal donor mass compatible with eventual crossing of the disc is found to be $0.575~\text{M}_\odot$.
Notably, the lower ejection velocity implied by this data set calls into question, assuming a current mass $0.3~\text{M}_\odot$ for US 708, ejection in a Chandrasekhar-mass SN, instead pointing to a terminal accretor mass of between $1.1~\text{M}_\odot$ and $1.2~\text{M}_\odot$ (Fig.~\ref{fig:US708ej} B and C). This, notably, would imply a SN involving the DDet mechanism (see Sec.~\ref{ssec:sub_ch}) and fall somewhere into the parameter space investigated by \cite{N1982b,N1982a}, \cite{WK2011} and \cite{NYL2017}. Assuming ejection in a Chandrasekhar mass SN, the most likely current mass for US 708 is in the range $0.34~\text{M}_\odot < M_\text{US708} < 0.37~\text{M}_\odot$, significantly higher than the assumed mass of $0.3~\text{M}_\odot$.
As in the 2015 data set, much rests on the question of whether US 708 is currently burning helium. If its mass is found to be below $0.3~\text{M}_\odot$, then, discounting errors, a Chandrasekhar-mass detonation would be conclusively ruled out.
As in the 2015 data set, as seen in Fig.~\ref{fig:hsearch}, a runaway mass greater than $0.4~\text{M}_\odot$ is ruled out for all but significantly super-Chandrasekhar-mass SNe.
However, it should be emphasized that the analytical power of correlating ejection velocity spectra and kinematic analysis would be greatly improved if done with distance measurements independent of stellar luminosity and mass estimates, i.e. parallax distances.

\section{Discussion} \label{sec:discussion}
Investigations of the runaway velocity of the surviving companions of WDs undergoing thermonuclear SNe are hampered by the unresolved nature of the most likely explosion mechanism. This is usually accompanied by the physics involved in the preceding evolution of the WD undergoing mass accretion being less than certain as well. Less uncertainty is involved in the evolution of the donor star, which can therefore serve as a convenient entry point for the modeling of ejection velocity spectra. 
Some of the simplifications employed in this study come with a number of caveats. While it was shown here that the initial orbital separation of the progenitor binary has little effect on the expected ejection velocity, previous studies \citep[e.g.][]{YL2004a,HK2004,NYL2017} suggest that the idiosyncratic evolution of the mass transfer rate associated with certain initial orbital separations do impact the ignition behavior of the accreted material on the WD. Besides calling into question the assumption of conservative mass transfer, as weak helium ignitions may lead to nova-like events, expelling part of the accumulated helium layer from the system, the ability of the accreting WD to accept additional material without triggering a supernova explosion will, in reality, be limited \citep[see e.g.][]{N1982b,N1982a,YL2004a,NYL2016,NYL2019}. The latter condition can realistically be expected to limit the ejection velocity spectra to those terminal accretor masses compatible with the assumed explosion mechanisms. The impact of non-conservative mass transfer, however, is less straightforward. As mass is lost from the system, compared to the conservative case, the metal content of the donor star will be higher for any combination of donor and accretor mass. As metallicity impacts the radius of the donor star (compare Fig.~\ref{fig:initpars} B), the corresponding orbital, and hence ejection velocity, can be expected to be lower. 
The results of these calculations could also be impacted by the effects of tides, especially heating effects, affecting the radius of the donor star, again leading to lower ejection velocities for the same mass combinations \citep{AP1987}.
The results presented here agree well with those of \cite{EF1990}, \cite{NYL2016} and \cite{NYL2019}. However, \cite{WH2009}, using population synthesis to study essentially the same problem, but limiting themselves to $M_\text{d,f}>0.6~\text{M}_\odot$, seem to find less strongly constrained ejection spectra. However, as they do not present ejection velocities in relation to terminal accretor mass, it is difficult to pinpoint the reason for this discrepancy. The most likely reason is the inclusion of bifurcations in the ejection velocity spectra, leading to a larger spread, as seen in Fig.~\ref{fig:v_bmass} (A). They further find relatively lower ejection velocities for $M_\text{d,f}\sim0.6~\text{M}_\odot$ and higher than found possible for $M_\text{d,f}\gtrsim1.0~\text{M}_\odot$ in this study. As their calculations include the evolution of the primordial main sequence binary, a systematic correlation between $M_\text{d,init}$ and $M_\text{WD,init}$ may be introduced which is absent from the models in this paper. This may explain the discrepancy at lower $M_\text{d,f}$. The discrepancy at higher $M_\text{d,f}$ may be explained by differences in metallicity and rotation, both absent from this study.
This applicability of the results obtained to observed runaway stars has been discussed at length. With respect to possible progenitor systems, the following objects can be commented upon:
\cite{VKT2012} and \cite{GMW2013}, independently showed $\text{CD}-30^\circ11223$ to contain a WD with a mass of $0.75-0.77~\text{M}_\odot$ and a hot subdwarf (sdB) of $0.44-0.48~\text{M}_\odot$ in a detached configuration and a period of $P = 0.04897906 \pm 0.00000004 \text{d}$. After entering a semidetached state, the WD in this system would need to accrete a substantial amount of He from its companion in order to become capable of producing an SN.
Interesting in the context of this study is V445 Pup, a nova-like variable that erupted in late 2000. \cite{AB2003} argued this object to represent a helium nova event with a WD accretor of $1.35~\text{M}_\odot$ \cite{KH2008} and a relatively massive $1.2-1.3~\text{M}_\odot$ helium star companion \citep{WSK2009}. It is currently unknown whether the donor star in this system is a giant or a HeMS star. If the donor can be shown to be a giant, and an SN eventually occurs, then runaway velocities would be too low to produce a HVS. If the donor is a HeMS star, then the fact that it is currently undergoing RLOF, indicates that it would have to have filled its Roche lobe immediately after the end of the most recent CE phase. Further considering the mass of the donor, ejection of a HVS is still unlikely if significant mass cannot be ejected from the system prior to the (assumed) detonation of the accretor.
Very recently, the discovery of a very short period binary composed of a $0.337 \pm 0.015~\text{M}_\odot$ helium sdOB and a $0.545 \pm 0.020~\text{M}_\odot$ WD was reported by \cite{KBM2020}. This system, if able to produce an SN at all, is unlikely to be able to produce a HVS.

Further theoretical exploration of the parameter space should include explicit treatments of the effects of non-conservative mass transfer and initial metallicity. Variations of the initial orbital separation at low total binary mass should also be considered. The effects of tidal interaction are likely important as well.

\section{Summary and conclusions} \label{sec:conclusions}
This paper presents a thorough study of the ejection velocity spectra for runaway stars resulting from thermonuclear SNe in the single degenerate helium donor channel. 

It is seen that the structural behavior of the donor star implies the existence of a maximum ejection velocity, correlated with the terminal mass of the donor star. The location, albeit not the value, of this maximum, is largely independent of the terminal mass of the accretor, and lies in the range $0.19~\text{M}_\odot < M_d < 0.25~\text{M}_\odot$. The value of the maximum ejection velocity, on the other hand, is dictated by the terminal mass of the accreting companion, with values correlated with higher terminal masses. It is found that maximum ejection velocities in excess of $1000~\text{km\,s}^{-1}$ can be attained with terminal accretor masses higher than $1.1~\text{M}_\odot$. 
This suggests that the SN ejection scenario is able to account for the existence of objects like the hypervelocity runaway sdO US 708 without a need for additional acceleration mechanisms, such as shock interaction. Concurrently, the assumed mass, $M_\text{US708}=0.3~\text{M}_\odot$, and inferred ejection velocity of this object in the 2015 data set is most consistent with a Chandrasekhar mass detonation, while more recently obtained proper motions suggest a sub-Chandrasekhar detonation with a terminal mass in the range $1.1~\text{M}_\odot$ to $1.2~\text{M}_\odot$. This result implies that the ejection of US 708 is compatible with its progenitor being identified as a single degenerate, helium accreting WD undergoing a SN according to the double detonation mechanism as proposed by \cite{N1982b,N1982a}.
Assuming a Chandrasekhar mass detonation with the 2020 data set would imply the current mass of US 708 to lie in the range $0.34~\text{M}_\odot < M_\text{US708} < 0.37~\text{M}_\odot$.

The maximum itself is a result of the mass transfer rate in the system exceeding the critical rate allowing for a decrease in orbital separation due to emission of GWR and transfer of orbital angular momentum due to mass transfer.
Since the maxima of the ejection velocity spectra are associated with terminal donor star masses below $0.3~\text{M}_\odot$, i.e. below the limit required for a star to retain the ability to burn helium, ejection at the highest velocities suggests that these objects are structurally different from slower ones and may result in the eventual production of single hypervelocity ELM WDs. This would be significant as ELM occurrence is currently thought to be dominated by binaries.
Production of a runaway with a mass higher than $0.3~\text{M}_\odot$ and exceeding $1000~\text{km\,s}^{-1}$ requires a terminal accretor mass in excess of the Chandrasekhar mass. Therefore, observation of a hypervelocity hot subdwarf with $v_\text{ej}>1000~\text{km\,s}^{-1}$ in its core helium burning phase would conclusively rule out sub-Chandrasekhar mass ejection in the SN scenario for that star.
In the same vein, observation of objects moving faster than the ejection velocity maximum will conclusively rule out SD SN ejection with the respective accretor mass.

While practicable studies of HVS ejection velocities for the purposes of the investigation of SD SN explosion mechanisms will require both more accurate observations and larger sample sizes than currently available, as demonstrated, they have the potential to provide constraints on terminal SN progenitor states not accessible to current observational techniques. In the future, data provided by large scale astrometry experiments such as Gaia \citep{GAIA-DR2} and the upcoming 4MOST instrument \citep{4MOST1} may be able to provide a sufficient number of observations.

\begin{acknowledgements}
Assistance of the Science and Technologies Facilities Council (UK), grant No. ST/S000453/1, is gratefully acknowledged. The author would like to thank Sergei Nayakshin, Phillip Podsiadlowski, Stephen Justham and Stephan Geier for useful discussions and Ulrich Heber for constructive suggestions which helped improve the paper.
\end{acknowledgements}

\bibliographystyle{aa}
\bibliography{paper.bib}{}

\begin{thebibliography}{96}
\expandafter\ifx\csname natexlab\endcsname\relax\def\natexlab#1{#1}\fi

\bibitem[{{Abadi} {et~al.}(2009){Abadi}, {Navarro}, \& {Steinmetz}}]{ANS2009}
{Abadi}, M.~G., {Navarro}, J.~F., \& {Steinmetz}, M. 2009, \apjl, 691, L63

\bibitem[{{Allen} \& {Santillan}(1991)}]{AS1991}
{Allen}, C. \& {Santillan}, A. 1991, \rmxaa, 22, 255

\bibitem[{{Applegate} \& {Patterson}(1987)}]{AP1987}
{Applegate}, J.~H. \& {Patterson}, J. 1987, \apjl, 322, L99

\bibitem[{{Arnett}(1969)}]{A1969}
{Arnett}, W.~D. 1969, \apss, 5, 180

\bibitem[{{Arnett}(1971)}]{A1971}
{Arnett}, W.~D. 1971, \apj, 169, 113

\bibitem[{{Ashok} \& {Banerjee}(2003)}]{AB2003}
{Ashok}, N.~M. \& {Banerjee}, D.~P.~K. 2003, \aap, 409, 1007

\bibitem[{{Bauer} {et~al.}(2019){Bauer}, {White}, \& {Bildsten}}]{BWB2019}
{Bauer}, E.~B., {White}, C.~J., \& {Bildsten}, L. 2019, arXiv e-prints,
  arXiv:1906.08941

\bibitem[{{Blaauw}(1961)}]{B1961}
{Blaauw}, A. 1961, Bulletin of the Astronomical Institutes of the Netherlands,
  15, 265

\bibitem[{{Bromley} {et~al.}(2006){Bromley}, {Kenyon}, {Geller}, {Barcikowski},
  {Brown}, \& {Kurtz}}]{BGB2006}
{Bromley}, B.~C., {Kenyon}, S.~J., {Geller}, M.~J., {et~al.} 2006, \apj, 653,
  1194

\bibitem[{{Brooks} {et~al.}(2017){Brooks}, {Schwab}, {Bildsten}, {Quataert}, \&
  {Paxton}}]{BSB2017}
{Brooks}, J., {Schwab}, J., {Bildsten}, L., {Quataert}, E., \& {Paxton}, B.
  2017, \apj, 843, 151

\bibitem[{{Brown}(2015)}]{B2015}
{Brown}, W.~R. 2015, \araa, 53, 15

\bibitem[{{Brown} {et~al.}(2005){Brown}, {Geller}, {Kenyon}, \&
  {Kurtz}}]{BGK2005}
{Brown}, W.~R., {Geller}, M.~J., {Kenyon}, S.~J., \& {Kurtz}, M.~J. 2005,
  \apjl, 622, L33

\bibitem[{{Brown} {et~al.}(2007){Brown}, {Geller}, {Kenyon}, {Kurtz}, \&
  {Bromley}}]{BGK2007}
{Brown}, W.~R., {Geller}, M.~J., {Kenyon}, S.~J., {Kurtz}, M.~J., \& {Bromley},
  B.~C. 2007, \apj, 671, 1708

\bibitem[{{Brown} {et~al.}(2020){Brown}, {Kilic}, {Kosakowski}, {Andrews},
  {Heinke}, {Ag{\"u}eros}, {Camilo}, {Gianninas}, {Hermes}, \&
  {Kenyon}}]{BKK2020}
{Brown}, W.~R., {Kilic}, M., {Kosakowski}, A., {et~al.} 2020, \apj, 889, 49

\bibitem[{{de Jong} {et~al.}(2016){de Jong}, {Barden}, {Bellido-Tirado},
  {Brynnel}, {Frey}, {Giannone}, {Haynes}, {Johl}, {Phillips}, {Schnurr},
  {Walcher}, {Winkler}, {Ansorge}, {Feltzing}, {McMahon}, {Baker}, {Caillier},
  {Dwelly}, {Gaessler}, {Iwert}, {Mandel}, {Piskunov}, {Pragt}, {Walton},
  {Bensby}, {Bergemann}, {Chiappini}, {Christlieb}, {Cioni}, {Driver},
  {Finoguenov}, {Helmi}, {Irwin}, {Kitaura}, {Kneib}, {Liske}, {Merloni},
  {Minchev}, {Richard}, \& {Starkenburg}}]{4MOST1}
{de Jong}, R.~S., {Barden}, S.~C., {Bellido-Tirado}, O., {et~al.} 2016, in
  \procspie, Vol. 9908, Ground-based and Airborne Instrumentation for Astronomy
  VI, 99081O

\bibitem[{{de la Fuente Marcos} \& {de la Fuente Marcos}(2019)}]{FF2019}
{de la Fuente Marcos}, R. \& {de la Fuente Marcos}, C. 2019, arXiv e-prints,
  arXiv:1906.05227

\bibitem[{{Di Stefano} {et~al.}(2011){Di Stefano}, {Voss}, \&
  {Claeys}}]{DVC2011}
{Di Stefano}, R., {Voss}, R., \& {Claeys}, J.~S.~W. 2011, The Astrophysical
  Journal, 738, L1

\bibitem[{{Edelmann} {et~al.}(2005){Edelmann}, {Napiwotzki}, {Heber},
  {Christlieb}, \& {Reimers}}]{ENH2005}
{Edelmann}, H., {Napiwotzki}, R., {Heber}, U., {Christlieb}, N., \& {Reimers},
  D. 2005, \apjl, 634, L181

\bibitem[{{Eggleton}(2011)}]{E2011B}
{Eggleton}, P. 2011, {Evolutionary Processes in Binary and Multiple Stars}
  (Cambridge Univ. Press)

\bibitem[{{Eggleton}(1983)}]{E1983}
{Eggleton}, P.~P. 1983, \apj, 268, 368

\bibitem[{{Ergma} \& {Fedorova}(1990)}]{EF1990}
{Ergma}, E.~V. \& {Fedorova}, A.~V. 1990, \apss, 163, 143

\bibitem[{{Erkal} {et~al.}(2019){Erkal}, {Boubert}, {Gualandris}, {Evans}, \&
  {Antonini}}]{EBG2019}
{Erkal}, D., {Boubert}, D., {Gualandris}, A., {Evans}, N.~W., \& {Antonini}, F.
  2019, \mnras, 483, 2007

\bibitem[{{Fink} {et~al.}(2007){Fink}, {Hillebrandt}, \& {R{\"o}pke}}]{FHR2007}
{Fink}, M., {Hillebrandt}, W., \& {R{\"o}pke}, F.~K. 2007, \aap, 476, 1133

\bibitem[{{Gaia Collaboration} {et~al.}(2018){Gaia Collaboration}, {Brown},
  {Vallenari}, {Prusti}, {de Bruijne}, {Babusiaux}, {Bailer-Jones}, {Biermann},
  {Evans}, {Eyer}, {Jansen}, {Jordi}, {Klioner}, {Lammers}, {Lindegren},
  {Luri}, {Mignard}, {Panem}, {Pourbaix}, {Randich}, {Sartoretti}, {Siddiqui},
  {Soubiran}, {van Leeuwen}, {Walton}, {Arenou}, {Bastian}, {Cropper},
  {Drimmel}, {Katz}, {Lattanzi}, {Bakker}, {Cacciari}, {Casta{\~n}eda},
  {Chaoul}, {Cheek}, {De Angeli}, {Fabricius}, {Guerra}, {Holl}, {Masana},
  {Messineo}, {Mowlavi}, {Nienartowicz}, {Panuzzo}, {Portell}, {Riello},
  {Seabroke}, {Tanga}, {Th{\'e}venin}, {Gracia-Abril}, {Comoretto},
  {Garcia-Reinaldos}, {Teyssier}, {Altmann}, {Andrae}, {Audard},
  {Bellas-Velidis}, {Benson}, {Berthier}, {Blomme}, {Burgess}, {Busso},
  {Carry}, {Cellino}, {Clementini}, {Clotet}, {Creevey}, {Davidson}, {De
  Ridder}, {Delchambre}, {Dell'Oro}, {Ducourant},
  {Fern{\'a}ndez-Hern{\'a}ndez}, {Fouesneau}, {Fr{\'e}mat}, {Galluccio},
  {Garc{\'\i}a-Torres}, {Gonz{\'a}lez-N{\'u}{\~n}ez}, {Gonz{\'a}lez-Vidal},
  {Gosset}, {Guy}, {Halbwachs}, {Hambly}, {Harrison}, {Hern{\'a}ndez},
  {Hestroffer}, {Hodgkin}, {Hutton}, {Jasniewicz}, {Jean-Antoine-Piccolo},
  {Jordan}, {Korn}, {Krone-Martins}, {Lanzafame}, {Lebzelter}, {L{\"o}ffler},
  {Manteiga}, {Marrese}, {Mart{\'\i}n-Fleitas}, {Moitinho}, {Mora}, {Muinonen},
  {Osinde}, {Pancino}, {Pauwels}, {Petit}, {Recio-Blanco}, {Richards},
  {Rimoldini}, {Robin}, {Sarro}, {Siopis}, {Smith}, {Sozzetti}, {S{\"u}veges},
  {Torra}, {van Reeven}, {Abbas}, {Abreu Aramburu}, {Accart}, {Aerts},
  {Altavilla}, {{\'A}lvarez}, {Alvarez}, {Alves}, {Anderson}, {Andrei},
  {Anglada Varela}, {Antiche}, {Antoja}, {Arcay}, {Astraatmadja}, {Bach},
  {Baker}, {Balaguer-N{\'u}{\~n}ez}, {Balm}, {Barache}, {Barata}, {Barbato},
  {Barblan}, {Barklem}, {Barrado}, {Barros}, {Barstow}, {Bartholom{\'e}
  Mu{\~n}oz}, {Bassilana}, {Becciani}, {Bellazzini}, {Berihuete}, {Bertone},
  {Bianchi}, {Bienaym{\'e}}, {Blanco-Cuaresma}, {Boch}, {Boeche}, {Bombrun},
  {Borrachero}, {Bossini}, {Bouquillon}, {Bourda}, {Bragaglia}, {Bramante},
  {Breddels}, {Bressan}, {Brouillet}, {Br{\"u}semeister}, {Brugaletta},
  {Bucciarelli}, {Burlacu}, {Busonero}, {Butkevich}, {Buzzi}, {Caffau},
  {Cancelliere}, {Cannizzaro}, {Cantat-Gaudin}, {Carballo}, {Carlucci},
  {Carrasco}, {Casamiquela}, {Castellani}, {Castro-Ginard}, {Charlot},
  {Chemin}, {Chiavassa}, {Cocozza}, {Costigan}, {Cowell}, {Crifo}, {Crosta},
  {Crowley}, {Cuypers}, {Dafonte}, {Damerdji}, {Dapergolas}, {David}, {David},
  {de Laverny}, {De Luise}, {De March}, {de Martino}, {de Souza}, {de Torres},
  {Debosscher}, {del Pozo}, {Delbo}, {Delgado}, {Delgado}, {Di Matteo},
  {Diakite}, {Diener}, {Distefano}, {Dolding}, {Drazinos}, {Dur{\'a}n},
  {Edvardsson}, {Enke}, {Eriksson}, {Esquej}, {Eynard Bontemps}, {Fabre},
  {Fabrizio}, {Faigler}, {Falc{\~a}o}, {Farr{\`a}s Casas}, {Federici},
  {Fedorets}, {Fernique}, {Figueras}, {Filippi}, {Findeisen}, {Fonti},
  {Fraile}, {Fraser}, {Fr{\'e}zouls}, {Gai}, {Galleti}, {Garabato},
  {Garc{\'\i}a-Sedano}, {Garofalo}, {Garralda}, {Gavel}, {Gavras}, {Gerssen},
  {Geyer}, {Giacobbe}, {Gilmore}, {Girona}, {Giuffrida}, {Glass}, {Gomes},
  {Granvik}, {Gueguen}, {Guerrier}, {Guiraud}, {Guti{\'e}rrez-S{\'a}nchez},
  {Haigron}, {Hatzidimitriou}, {Hauser}, {Haywood}, {Heiter}, {Helmi}, {Heu},
  {Hilger}, {Hobbs}, {Hofmann}, {Holland}, {Huckle}, {Hypki}, {Icardi},
  {Jan{\ss}en}, {Jevardat de Fombelle}, {Jonker}, {Juh{\'a}sz}, {Julbe},
  {Karampelas}, {Kewley}, {Klar}, {Kochoska}, {Kohley}, {Kolenberg},
  {Kontizas}, {Kontizas}, {Koposov}, {Kordopatis}, {Kostrzewa-Rutkowska},
  {Koubsky}, {Lambert}, {Lanza}, {Lasne}, {Lavigne}, {Le Fustec}, {Le
  Poncin-Lafitte}, {Lebreton}, {Leccia}, {Leclerc}, {Lecoeur-Taibi},
  {Lenhardt}, {Leroux}, {Liao}, {Licata}, {Lindstr{\o}m}, {Lister}, {Livanou},
  {Lobel}, {L{\'o}pez}, {Managau}, {Mann}, {Mantelet}, {Marchal}, {Marchant},
  {Marconi}, {Marinoni}, {Marschalk{\'o}}, {Marshall}, {Martino}, {Marton},
  {Mary}, {Massari}, {Matijevi{\v{c}}}, {Mazeh}, {McMillan}, {Messina},
  {Michalik}, {Millar}, {Molina}, {Molinaro}, {Moln{\'a}r}, {Montegriffo},
  {Mor}, {Morbidelli}, {Morel}, {Morris}, {Mulone}, {Muraveva}, {Musella},
  {Nelemans}, {Nicastro}, {Noval}, {O'Mullane}, {Ord{\'e}novic},
  {Ord{\'o}{\~n}ez-Blanco}, {Osborne}, {Pagani}, {Pagano}, {Pailler},
  {Palacin}, {Palaversa}, {Panahi}, {Pawlak}, {Piersimoni}, {Pineau}, {Plachy},
  {Plum}, {Poggio}, {Poujoulet}, {Pr{\v{s}}a}, {Pulone}, {Racero}, {Ragaini},
  {Rambaux}, {Ramos-Lerate}, {Regibo}, {Reyl{\'e}}, {Riclet}, {Ripepi}, {Riva},
  {Rivard}, {Rixon}, {Roegiers}, {Roelens}, {Romero-G{\'o}mez}, {Rowell},
  {Royer}, {Ruiz-Dern}, {Sadowski}, {Sagrist{\`a} Sell{\'e}s}, {Sahlmann},
  {Salgado}, {Salguero}, {Sanna}, {Santana-Ros}, {Sarasso}, {Savietto},
  {Schultheis}, {Sciacca}, {Segol}, {Segovia}, {S{\'e}gransan}, {Shih},
  {Siltala}, {Silva}, {Smart}, {Smith}, {Solano}, {Solitro}, {Sordo}, {Soria
  Nieto}, {Souchay}, {Spagna}, {Spoto}, {Stampa}, {Steele},
  {Steidelm{\"u}ller}, {Stephenson}, {Stoev}, {Suess}, {Surdej}, {Szabados},
  {Szegedi-Elek}, {Tapiador}, {Taris}, {Tauran}, {Taylor}, {Teixeira},
  {Terrett}, {Teyssand ier}, {Thuillot}, {Titarenko}, {Torra Clotet}, {Turon},
  {Ulla}, {Utrilla}, {Uzzi}, {Vaillant}, {Valentini}, {Valette}, {van Elteren},
  {Van Hemelryck}, {van Leeuwen}, {Vaschetto}, {Vecchiato}, {Veljanoski},
  {Viala}, {Vicente}, {Vogt}, {von Essen}, {Voss}, {Votruba}, {Voutsinas},
  {Walmsley}, {Weiler}, {Wertz}, {Wevers}, {Wyrzykowski}, {Yoldas},
  {{\v{Z}}erjal}, {Ziaeepour}, {Zorec}, {Zschocke}, {Zucker}, {Zurbach}, \&
  {Zwitter}}]{GAIA-DR2}
{Gaia Collaboration}, {Brown}, A.~G.~A., {Vallenari}, A., {et~al.} 2018, \aap,
  616, A1

\bibitem[{{Gamezo} {et~al.}(2005){Gamezo}, {Khokhlov}, \& {Oran}}]{GKO2005}
{Gamezo}, V.~N., {Khokhlov}, A.~M., \& {Oran}, E.~S. 2005, \apj, 623, 337

\bibitem[{{Geier} {et~al.}(2015){Geier}, {F{\"u}rst}, {Ziegerer}, {Kupfer},
  {Heber}, {Irrgang}, {Wang}, {Liu}, {Han}, {Sesar}, {Levitan}, {Kotak},
  {Magnier}, {Smith}, {Burgett}, {Chambers}, {Flewelling}, {Kaiser},
  {Wainscoat}, \& {Waters}}]{GFZ2015}
{Geier}, S., {F{\"u}rst}, F., {Ziegerer}, E., {et~al.} 2015, Science, 347, 1126

\bibitem[{{Geier} {et~al.}(2013){Geier}, {Marsh}, {Wang}, {Dunlap}, {Barlow},
  {Schaffenroth}, {Chen}, {Irrgang}, {Maxted}, {Ziegerer}, {Kupfer},
  {Miszalski}, {Heber}, {Han}, {Shporer}, {Telting}, {G{\"a}nsicke},
  {{\O}stensen}, {O'Toole}, \& {Napiwotzki}}]{GMW2013}
{Geier}, S., {Marsh}, T.~R., {Wang}, B., {et~al.} 2013, \aap, 554, A54

\bibitem[{{Hachisu} {et~al.}(2012){Hachisu}, {Kato}, {Saio}, \&
  {Nomoto}}]{HKSN2012}
{Hachisu}, I., {Kato}, M., {Saio}, H., \& {Nomoto}, K. 2012, \apj, 744, 69

\bibitem[{{Han}(2008)}]{H2008}
{Han}, Z. 2008, \apjl, 677, L109

\bibitem[{{Hansen} \& {Wheeler}(1969)}]{HW1969}
{Hansen}, C.~J. \& {Wheeler}, J.~C. 1969, \apss, 3, 464

\bibitem[{{Heber}(2009)}]{H2009}
{Heber}, U. 2009, \araa, 47, 211

\bibitem[{{Heber}(2016)}]{H2016}
{Heber}, U. 2016, \pasp, 128, 082001

\bibitem[{{Hillebrandt} \& {Niemeyer}(2000)}]{HN2000}
{Hillebrandt}, W. \& {Niemeyer}, J.~C. 2000, \araa, 38, 191

\bibitem[{{Hills}(1988)}]{H1988}
{Hills}, J.~G. 1988, \nat, 331, 687

\bibitem[{{Hills}(1991)}]{H1991}
{Hills}, J.~G. 1991, \aj, 102, 704

\bibitem[{{Hills}(1992)}]{H1992}
{Hills}, J.~G. 1992, \aj, 103, 1955

\bibitem[{{Hirsch} {et~al.}(2005){Hirsch}, {Heber}, {O'Toole}, \&
  {Bresolin}}]{HHOB2005}
{Hirsch}, H.~A., {Heber}, U., {O'Toole}, S.~J., \& {Bresolin}, F. 2005, \aap,
  444, L61

\bibitem[{{Hoogerwerf} {et~al.}(2001){Hoogerwerf}, {de Bruijne}, \& {de
  Zeeuw}}]{HBZ2001}
{Hoogerwerf}, R., {de Bruijne}, J.~H.~J., \& {de Zeeuw}, P.~T. 2001, \aap, 365,
  49

\bibitem[{{Howell} {et~al.}(2006){Howell}, {Sullivan}, {Nugent}, {Ellis},
  {Conley}, {Le Borgne}, {Carlberg}, {Guy}, {Balam}, \& {Basa}}]{HSN2006}
{Howell}, D.~A., {Sullivan}, M., {Nugent}, P.~E., {et~al.} 2006, \nat, 443, 308

\bibitem[{{Hoyle} \& {Fowler}(1960)}]{HF1960}
{Hoyle}, F. \& {Fowler}, W.~A. 1960, \apj, 132, 565

\bibitem[{{Hut}(1981)}]{H1981}
{Hut}, P. 1981, \aap, 99, 126

\bibitem[{{Irrgang} {et~al.}(2019){Irrgang}, {Geier}, {Heber}, {Kupfer}, \&
  {F{\"u}rst}}]{IGH2019}
{Irrgang}, A., {Geier}, S., {Heber}, U., {Kupfer}, T., \& {F{\"u}rst}, F. 2019,
  \aap, 628, L5

\bibitem[{{Irrgang} {et~al.}(2018){Irrgang}, {Kreuzer}, \& {Heber}}]{IKH2018}
{Irrgang}, A., {Kreuzer}, S., \& {Heber}, U. 2018, \aap, 620, A48

\bibitem[{{Irrgang} {et~al.}(2013){Irrgang}, {Wilcox}, {Tucker}, \&
  {Schiefelbein}}]{IWT2013}
{Irrgang}, A., {Wilcox}, B., {Tucker}, E., \& {Schiefelbein}, L. 2013, \aap,
  549, A137

\bibitem[{{Justham} {et~al.}(2009){Justham}, {Wolf}, {Podsiadlowski}, \&
  {Han}}]{JWP2009}
{Justham}, S., {Wolf}, C., {Podsiadlowski}, P., \& {Han}, Z. 2009, \aap, 493,
  1081

\bibitem[{Kato \& Hachisu(2004)}]{HK2004}
Kato, M. \& Hachisu, I. 2004, The Astrophysical Journal Letters, 613, L129

\bibitem[{{Kato} {et~al.}(2008){Kato}, {Hachisu}, \& {Kiyota}}]{KH2008}
{Kato}, M., {Hachisu}, I., \& {Kiyota}, S. 2008, in Astronomical Society of the
  Pacific Conference Series, Vol. 391, Hydrogen-Deficient Stars, ed.
  A.~{Werner} \& T.~{Rauch}, 267

\bibitem[{{Khokhlov} {et~al.}(1997){Khokhlov}, {Oran}, \& {Wheeler}}]{KOW1997}
{Khokhlov}, A.~M., {Oran}, E.~S., \& {Wheeler}, J.~C. 1997, \apj, 478, 678

\bibitem[{{Kippenhahn} {et~al.}(2012){Kippenhahn}, {Weigert}, \&
  {Weiss}}]{STELLAR_STRUCTURE_AND_EVOLUTION}
{Kippenhahn}, R., {Weigert}, A., \& {Weiss}, A. 2012, {Stellar Structure and
  Evolution}

\bibitem[{{Koposov} {et~al.}(2020){Koposov}, {Boubert}, {Li}, {Erkal}, {Da
  Costa}, {Zucker}, {Ji}, {Kuehn}, {Lewis}, {Mackey}, {Simpson}, {Shipp},
  {Wan}, {Belokurov}, {Bland-Hawthorn}, {Martell}, {Nordlander}, {Pace}, {De
  Silva}, {Wang}, \& {S5 collaboration}}]{KBD2020}
{Koposov}, S.~E., {Boubert}, D., {Li}, T.~S., {et~al.} 2020, \mnras, 491, 2465

\bibitem[{{Kupfer} {et~al.}(2020){Kupfer}, {Bauer}, {Marsh}, {van Roestel},
  {Bellm}, {Burdge}, {Coughlin}, {Fuller}, {Hermes}, {Bildsten}, {Kulkarni},
  {Prince}, {Szkody}, {Dhillon}, {Murawski}, {Burruss}, {Dekany}, {Delacroix},
  {Drake}, {Duev}, {Feeney}, {Graham}, {Kaplan}, {Laher}, {Littlefair},
  {Masci}, {Riddle}, {Rusholme}, {Serabyn}, {Smith}, {Shupe}, \&
  {Soumagnac}}]{KBM2020}
{Kupfer}, T., {Bauer}, E.~B., {Marsh}, T.~R., {et~al.} 2020, arXiv e-prints,
  arXiv:2002.01485

\bibitem[{Landau \& Livshitz(1975)}]{LL1975}
Landau, L. \& Livshitz, E. 1975, The Classical Theory of Fields, Course of
  theoretical physics (Butterworth Heinemann)

\bibitem[{{Liu} {et~al.}(2010){Liu}, {Chen}, {Wang}, \& {Han}}]{LCW2010}
{Liu}, W.~M., {Chen}, W.~C., {Wang}, B., \& {Han}, Z.~W. 2010, \aap, 523, A3

\bibitem[{{Livne} \& {Arnett}(1995)}]{LA1995}
{Livne}, E. \& {Arnett}, D. 1995, \apj, 452, 62

\bibitem[{{Mestel}(1968)}]{M1968}
{Mestel}, L. 1968, \mnras, 138, 359

\bibitem[{{Neunteufel} {et~al.}(2016){Neunteufel}, {Yoon}, \&
  {Langer}}]{NYL2016}
{Neunteufel}, P., {Yoon}, S.-C., \& {Langer}, N. 2016, \aap, 589, A43

\bibitem[{{Neunteufel} {et~al.}(2017){Neunteufel}, {Yoon}, \&
  {Langer}}]{NYL2017}
{Neunteufel}, P., {Yoon}, S.-C., \& {Langer}, N. 2017, \aap, 602, A55

\bibitem[{{Neunteufel} {et~al.}(2019){Neunteufel}, {Yoon}, \&
  {Langer}}]{NYL2019}
{Neunteufel}, P., {Yoon}, S.~C., \& {Langer}, N. 2019, \aap, 627, A14

\bibitem[{{Nomoto}(1980)}]{N1980P}
{Nomoto}, K. 1980, in Texas Workshop on Type I Supernovae, ed. J.~C. {Wheeler},
  164--181

\bibitem[{{Nomoto}(1982{\natexlab{a}})}]{N1982b}
{Nomoto}, K. 1982{\natexlab{a}}, \apj, 257, 780

\bibitem[{{Nomoto}(1982{\natexlab{b}})}]{N1982a}
{Nomoto}, K. 1982{\natexlab{b}}, \apj, 253, 798

\bibitem[{{Nomoto} {et~al.}(1984){Nomoto}, {Thielemann}, \& {Yokoi}}]{NTY1984}
{Nomoto}, K., {Thielemann}, F.~K., \& {Yokoi}, K. 1984, \apj, 286, 644

\bibitem[{{O'Leary} \& {Loeb}(2008)}]{LL2008}
{O'Leary}, R.~M. \& {Loeb}, A. 2008, \mnras, 383, 86

\bibitem[{{Pakmor} {et~al.}(2013){Pakmor}, {Kromer}, {Taubenberger}, \&
  {Springel}}]{PKT2013}
{Pakmor}, R., {Kromer}, M., {Taubenberger}, S., \& {Springel}, V. 2013, \apjl,
  770, L8

\bibitem[{{Paxton} {et~al.}(2011){Paxton}, {Bildsten}, {Dotter}, {Herwig},
  {Lesaffre}, \& {Timmes}}]{MESA1}
{Paxton}, B., {Bildsten}, L., {Dotter}, A., {et~al.} 2011, \apjs, 192, 3

\bibitem[{{Paxton} {et~al.}(2013){Paxton}, {Cantiello}, {Arras}, {Bildsten},
  {Brown}, {Dotter}, {Mankovich}, {Montgomery}, {Stello}, {Timmes}, \&
  {Townsend}}]{MESA2}
{Paxton}, B., {Cantiello}, M., {Arras}, P., {et~al.} 2013, \apjs, 208, 4

\bibitem[{{Paxton} {et~al.}(2015){Paxton}, {Marchant}, {Schwab}, {Bauer},
  {Bildsten}, {Cantiello}, {Dessart}, {Farmer}, {Hu}, {Langer}, {Townsend},
  {Townsley}, \& {Timmes}}]{MESA3}
{Paxton}, B., {Marchant}, P., {Schwab}, J., {et~al.} 2015, \apjs, 220, 15

\bibitem[{{Paxton} {et~al.}(2018){Paxton}, {Schwab}, {Bauer}, {Bildsten},
  {Blinnikov}, {Duffell}, {Farmer}, {Goldberg}, {Marchant}, {Sorokina},
  {Thoul}, {Townsend}, \& {Timmes}}]{MESA4}
{Paxton}, B., {Schwab}, J., {Bauer}, E.~B., {et~al.} 2018, \apjs, 234, 34

\bibitem[{{Paxton} {et~al.}(2019){Paxton}, {Smolec}, {Schwab}, {Gautschy},
  {Bildsten}, {Cantiello}, {Dotter}, {Farmer}, {Goldberg}, {Jermyn}, {Kanbur},
  {Marchant}, {Thoul}, {Townsend}, {Wolf}, {Zhang}, \& {Timmes}}]{MESA5}
{Paxton}, B., {Smolec}, R., {Schwab}, J., {et~al.} 2019, \apjs, 243, 10

\bibitem[{{Phillips}(1993)}]{P1993}
{Phillips}, M.~M. 1993, \apjl, 413, L105

\bibitem[{{Piffl} {et~al.}(2014){Piffl}, {Scannapieco}, {Binney}, {Steinmetz},
  {Scholz}, {Williams}, {de Jong}, {Kordopatis}, {Matijevi{\v c}},
  {Bienaym{\'e}}, {Bland-Hawthorn}, {Boeche}, {Freeman}, {Gibson}, {Gilmore},
  {Grebel}, {Helmi}, {Munari}, {Navarro}, {Parker}, {Reid}, {Seabroke},
  {Watson}, {Wyse}, \& {Zwitter}}]{PSB2014}
{Piffl}, T., {Scannapieco}, C., {Binney}, J., {et~al.} 2014, \aap, 562, A91

\bibitem[{{Plewa} {et~al.}(2004){Plewa}, {Calder}, \& {Lamb}}]{PCL2004}
{Plewa}, T., {Calder}, A.~C., \& {Lamb}, D.~Q. 2004, \apjl, 612, L37

\bibitem[{{Press} {et~al.}(1992){Press}, {Teukolsky}, {Vetterling}, \&
  {Flannery}}]{RECIPES_C}
{Press}, W.~H., {Teukolsky}, S.~A., {Vetterling}, W.~T., \& {Flannery}, B.~P.
  1992, {Numerical recipes in C. The art of scientific computing}

\bibitem[{{Raddi} {et~al.}(2019){Raddi}, {Hollands}, {Koester}, {Hermes},
  {G{\"a}nsicke}, {Heber}, {Shen}, {Townsley}, {Pala}, {Reding}, {Toloza},
  {Pelisoli}, {Geier}, {Gentile Fusillo}, {Munari}, \& {Strader}}]{RHK2019}
{Raddi}, R., {Hollands}, M.~A., {Koester}, D., {et~al.} 2019, \mnras, 489, 1489

\bibitem[{{Ritter}(1988)}]{R1988}
{Ritter}, H. 1988, \aap, 202, 93

\bibitem[{{Scholz}(2018)}]{S2018}
{Scholz}, R.-D. 2018, Research Notes of the American Astronomical Society, 2,
  211

\bibitem[{{Sesana} {et~al.}(2009){Sesana}, {Madau}, \& {Haardt}}]{SMH2009}
{Sesana}, A., {Madau}, P., \& {Haardt}, F. 2009, \mnras, 392, L31

\bibitem[{{Shen} {et~al.}(2018){Shen}, {Boubert}, {G{\"a}nsicke}, {Jha},
  {Andrews}, {Chomiuk}, {Foley}, {Fraser}, {Gromadzki}, {Guillochon}, {Kotze},
  {Maguire}, {Siebert}, {Smith}, {Strader}, {Badenes}, {Kerzendorf}, {Koester},
  {Kromer}, {Miles}, {Pakmor}, {Schwab}, {Toloza}, {Toonen}, {Townsley}, \&
  {Williams}}]{SBG2018}
{Shen}, K.~J., {Boubert}, D., {G{\"a}nsicke}, B.~T., {et~al.} 2018, \apj, 865,
  15

\bibitem[{{Sim} {et~al.}(2010){Sim}, {R{\"o}pke}, {Hillebrandt}, {Kromer},
  {Pakmor}, {Fink}, {Ruiter}, \& {Seitenzahl}}]{SRH2010}
{Sim}, S.~A., {R{\"o}pke}, F.~K., {Hillebrandt}, W., {et~al.} 2010, \apjl, 714,
  L52

\bibitem[{{Soker}(2013)}]{S2013}
{Soker}, N. 2013, in IAU Symposium, Vol. 281, Binary Paths to Type Ia
  Supernovae Explosions, ed. R.~{Di Stefano}, M.~{Orio}, \& M.~{Moe}, 72--75

\bibitem[{{Tauris}(2015)}]{T2015T}
{Tauris}, T.~M. 2015, \mnras, 448, L6

\bibitem[{{Tutukov} \& {Yungelson}(1979)}]{TY1979}
{Tutukov}, A.~V. \& {Yungelson}, L.~R. 1979, \actaa, 29, 665

\bibitem[{{Vennes} {et~al.}(2012){Vennes}, {Kawka}, {O'Toole}, {N{\'e}meth}, \&
  {Burton}}]{VKT2012}
{Vennes}, S., {Kawka}, A., {O'Toole}, S.~J., {N{\'e}meth}, P., \& {Burton}, D.
  2012, \apjl, 759, L25

\bibitem[{{Vennes} {et~al.}(2017){Vennes}, {Nemeth}, {Kawka}, {Thorstensen},
  {Khalack}, {Ferrario}, \& {Alper}}]{VNK2017}
{Vennes}, S., {Nemeth}, P., {Kawka}, A., {et~al.} 2017, Science, 357, 680

\bibitem[{{Wang} \& {Han}(2009)}]{WH2009}
{Wang}, B. \& {Han}, Z. 2009, \aap, 508, L27

\bibitem[{{Wang} {et~al.}(2013){Wang}, {Justham}, \& {Han}}]{WJH2013}
{Wang}, B., {Justham}, S., \& {Han}, Z. 2013, A\&A, 559, A94

\bibitem[{{Webbink}(1984)}]{W1984}
{Webbink}, R.~F. 1984, \apj, 277, 355

\bibitem[{{Woosley} \& {Kasen}(2011)}]{WK2011}
{Woosley}, S.~E. \& {Kasen}, D. 2011, \apj, 734, 38

\bibitem[{{Woosley} \& {Weaver}(1994)}]{WW1994}
{Woosley}, S.~E. \& {Weaver}, T.~A. 1994, \apj, 423, 371

\bibitem[{{Woudt} {et~al.}(2009){Woudt}, {Steeghs}, {Karovska}, {Warner},
  {Groot}, {Nelemans}, {Roelofs}, {Marsh}, {Nagayama}, {Smits}, \&
  {O'Brien}}]{WSK2009}
{Woudt}, P.~A., {Steeghs}, D., {Karovska}, M., {et~al.} 2009, \apj, 706, 738

\bibitem[{{Yoon} \& {Langer}(2004{\natexlab{a}})}]{YL2004b}
{Yoon}, S.-C. \& {Langer}, N. 2004{\natexlab{a}}, \aap, 419, 645

\bibitem[{{Yoon} \& {Langer}(2004{\natexlab{b}})}]{YL2004a}
{Yoon}, S.-C. \& {Langer}, N. 2004{\natexlab{b}}, \aap, 419, 623

\bibitem[{{Yoon} \& {Langer}(2005)}]{YL2005}
{Yoon}, S.-C. \& {Langer}, N. 2005, \aap, 435, 967

\bibitem[{Yu \& Tremaine(2003)}]{YT2003}
Yu, Q. \& Tremaine, S. 2003, The Astrophysical Journal, 599, 1129

\bibitem[{{Yungelson}(2008)}]{Yu2008}
{Yungelson}, L.~R. 2008, Astronomy Letters, 34, 620

\bibitem[{{Zhang} {et~al.}(2019){Zhang}, {Fuller}, {Schwab}, \&
  {Foley}}]{ZF2019}
{Zhang}, M., {Fuller}, J., {Schwab}, J., \& {Foley}, R.~J. 2019, \apj, 872, 29

\end{thebibliography}

\appendix

\section{Effects of initial orbital separation} \label{app:as}
As described in Sec.~\ref{sec:methods}, initial orbital separations $a_\text{init}(\xi)$ in this paper were chosen such that Eq.~\ref{eq:eggle} satisfies $R_\text{d,RL}=\xi \cdot R_\text{d}$ with $\xi=1.005$. Since this choice results in a variety of initial orbital separations and initial periods, orbital separations and initial periods will not be comparable between individual systems, even between systems with equal total or component masses. As argued, the window for RLOF during the HeMS of the donor star is quite narrow in terms of $\xi$ and the effect of a different choice of that parameter on the expected ejection velocity correspondingly small. In order to test this argument, the full grid was rerun with initial orbital separations corresponding to $\xi=1.01$. 
\begin{figure}
	\input{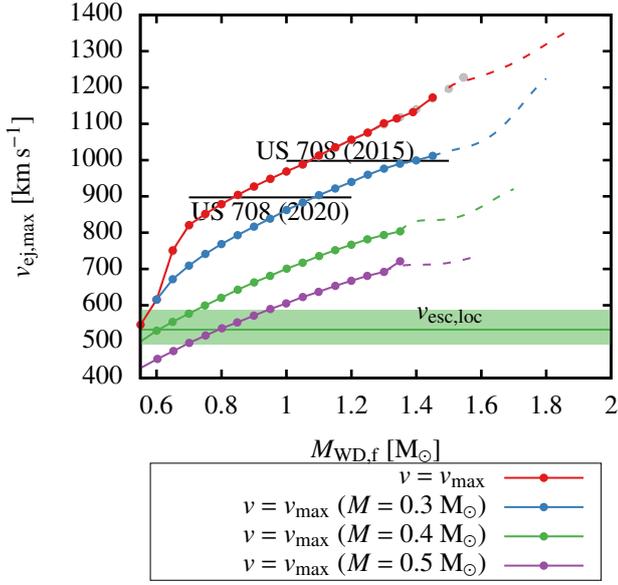}
	\caption{Like Fig.~\ref{fig:hsearch}, but $a_\text{init}(\xi = 1.01)$ (See Sec~\ref{sec:methods}). The gray line indicates $v_{max}$ as in Fig.~\ref{fig:hsearch} for comparison.}  \label{fig:hsearch2}
\end{figure}

As seen in Fig.~\ref{fig:hsearch2}, the maximum ejection velocity with $\xi=1.01$ is not significantly increased compared with $\xi=1.005$ at the same mass with analogous $M_\text{d,f}$. It can reasonably concluded that the initial orbital separation is of secondary importance for the question of ejection velocity maxima. However, due to the desire to access the entire mass spectrum with two sets of initial models, this increase is small. While the predictive power hat high values of $M_\text{d,init}$ can be called reasonable, more work is required in cases of low $M_\text{d,init}$.

\section{HVS kinematic analysis} \label{app:traj}
Trajectory and time-of-flight data as described in Sec.~\ref{sec:US708} were calculated by numerically solving the the Newtonian equations of motion in the well known form
\begin{equation}
m\frac{d}{dt} \frac{d\vec{x}}{dt} = - \nabla \Phi(\vec{x})
\end{equation}
with $\Phi$ the Galactic potential. The numerical solver utilized here was newly developed on the basis of a fourth-order Runge-Kutta integrator with adaptive step size control as described by \cite{RECIPES_C}.
The Galactic potential was assumed to be static and correspond to Model 1 put forward by \cite{IWT2013} as a revision of \cite{AS1991} in the form 
\begin{equation} \label{eq:model1.1}
	\Phi_b(R)= -\frac{M_b}{\sqrt{R^2+b_b^2}}
\end{equation}
for the bulge, where $R$ is the distance from the Galactic center
\begin{equation}
\Phi_d(r,z)= -\frac{M_d}{\sqrt{r^2+(a_d+\sqrt{z^2+b_d^2})}}
\end{equation}
for the disk, where $r$ is the distance from the Galactic center in the $x$-$y$-plane and $z$ is the distance from the $x$-$y$-plane and
\begin{align} \label{eq:model1.4}
\Phi_h(R)&= -\frac{M_h}{a_h}\left[ \frac{1}{\gamma -1} \ln\left( \frac{1+(R/a_h)^{\gamma-1}}{1+(\Lambda/a_h)^{\gamma-1}} \right) - \frac{(\Lambda/a_h)^{\gamma-1}}{(1 +\Lambda/a_h)^{\gamma-1}} \right]~\text{if}~R < \Lambda \nonumber \\
&= -\frac{M_h}{R} \frac{(\Lambda/a_h)^{\gamma-1}}{(1 +\Lambda/a_h)^{\gamma-1}}~\text{if}~R \geq \Lambda 
\end{align}

with $\gamma=2$ and the other parameters given in Tab.~\ref{tab:modelp}.
\begin{table} 
	\caption{Parameters used in Eqs.~\ref{eq:model1.1}-\ref{eq:model1.4}.} \label{tab:modelp}
\begin{tabular}{ccccc} 
	\hline 
	\hline
	& $M_\text{b/d/h}~[\text{M}_\text{G}]$ & $a_\text{d/h}~[\text{kpc}]$ & $b_\text{b/d/h}~[\text{kpc}]$ & $\Lambda~[\text{kpc}]$  \\  
	\hline
	Bulge$_\text{b}$& $409\pm63$  &  & $0.23\pm0.03$  &\\ 
	Disk$_\text{d}$ & $2856^{+376}_{-202}$  & $4.22^{+0.53}_{-0.99}$ & $0.292^{+0.020}_{-0.025}$ &\\ 
	Halo$_\text{h}$ & $1018^{+27933}_{-603}$ & $2.562^{+25.963}_{-1.419}$ & $4.22^{+0.53}_{-0.99}$ &$200^{+0}_{-82}$\\ 
	\hline 
\end{tabular} 
\end{table}

The distance of the Sun from the Galactic center is given as $r_\text{Sol} = 8.40\pm0.08~\text{kpc}$.

Using the trajectory solver with the parameters as described above, and letting the inferred mass of US 708 be a free parameter, as described in Sec.~\ref{sec:US708}, the time-of-flight (Fig.~\ref{fig:US708tof}) and ejection location (Fig.~\ref{fig:US708ejloc}) can be calculated. 
\begin{figure} 
	\input{US708tof}
	\caption{Mass-dependent time of flight (TOF) since ejection from the disc for US 708, once for proper motions as obtained by \cite{GFZ2015} and once for proper motions as published by GAIA DR2 \citep{GAIA-DR2}. The preferred model with $M_\text{US708}=0.3~M_\odot$ is highlighted in green. Note that the counter-intuitively shorter TOF at similar mass obtained for the slower space velocity of the 2020 data set is a result of the shorter distance traveled to the disc-crossing point. This shorter distance, in turn, is derived from the reoriented direction of travel as compared to the 2015 data set (See Fig.~\ref{fig:US708ejloc}).} \label{fig:US708tof}
\end{figure}
\begin{figure} 
	\input{US708ejloc}
	\input{US708ejloc-N}
	\caption{Current inferred location, depending on assumed mass ($M_\text{US708}$) of US 708, and inferred origins relative to the Galactic center and the position of the Sun in the Galactic x-y plane.
	Panel (A) uses proper motions as obtained by \cite{GFZ2015}, panel (B) uses proper motions obtained by GAIA \citep{GAIA-DR2} The location, origin and past trajectory of the preferred model with $M_\text{US708}=0.3~\text{M}_\odot$ is highlighted in green.} \label{fig:US708ejloc}
\end{figure}

Results for both time-of-flight and ejection location are in good agreement with the more detailed calculations presented by \cite{GFZ2015} for the preferred model of $M_\text{US708} = 0.3~\text{M}_\odot$.

\end{document}